\documentclass[aps, pre, reprint, showpacs,preprintnumbers, a4paper]{revtex4-1}
\usepackage[utf8x]{inputenc}
\usepackage[english]{babel}
\usepackage{graphicx,subfigure}
\usepackage{dcolumn}
\usepackage{amsmath}
\usepackage{amssymb}
\usepackage{bm}
\usepackage{color}     

\begin{document}
 
\title{Delayed feedback control of self-mobile cavity solitons in a wide-aperture laser with a saturable absorber}
\date{\today}
\author{T. Schemmelmann$^{1}$, F. Tabbert$^{1}$, A. Pimenov$^{2}$, A. G. Vladimirov$^{2}$, S. V. Gurevich$^{1,3}$}
\email{gurevics@uni-muenster.de}
\affiliation{$^{1}$Institute for Theoretical Physics, University of M\"unster, Wilhelm-Klemm-Str.9, D-48149, M\"unster, Germany}
\affiliation{$^{2}$Weierstrass Institute for Applied Analysis and Stochastics, Mohrenstrasse 39, D-10117 Berlin, Germany}
\affiliation{$^{3}$Center for Nonlinear Science (CeNoS), University of M\"unster, Corrensstr.\,2, 48149 M\"unster, Germany}

\begin{abstract}
We investigate the spatiotemporal dynamics of cavity solitons in a broad area vertical-cavity surface-emitting laser with saturable absorption subjected to time-delayed optical feedback. Using a combination of analytical, numerical and path continuation methods we analyze the bifurcation structure of stationary and moving cavity solitons and identify two different types of traveling localized solutions, corresponding to  slow and fast motion. We show that the delay impacts both stationary and moving solutions either causing drifting and wiggling dynamics of initially stationary cavity solitons or leading to stabilization of intrinsically moving solutions. Finally, we demonstrate that the fast cavity solitons can be associated with a lateral mode-locking regime in a broad-area laser with a single longitudinal mode.
\end{abstract}
\maketitle

\begin{section}{Introduction}
The formation of dissipative localized structures was reported in various research fields ranging from chemistry to ecology and social science~\cite{AA-LNP-08,PurwinsDS2010,  Lier2013,VegetationPRL2001,Mikhailov2006,Tlidi20140101,ShortCrime}. Localized structures of light in the transverse section of externally driven nonlinear resonators and lasers are often called cavity solitons (CSs)~\cite{L-CSF-94,Rosanov_po96,MT-JOSAB-04,Ackemann_aamoo09}. Since their experimental demonstration CSs in semiconductor cavities attracted growing interest due to potential applications for information storage and processing~\cite{Jacobo2012, PedaciAPL2008}. CSs usually appear as bright and/or dark spots in the transverse plane of nonlinear cavities, which is orthogonal to the propagation axis. Recently, Vertical-Cavity Surface-Emitting Lasers (VCSELs) attracted considerable interest in CSs studies and different experimental techniques of the CS generation in these lasers have been reported. This includes coherent optical injection in combination with a narrow writing beam, frequency selective feedback and saturable absorption (see~\cite{Barbay2011} for a review). In particular, VCSELs with saturable absorption are of interest since they do not require an external holding beam for the CS generation, which significantly simplifies the underlying system. Therefore, the formation of CSs in VCSELs with a saturable absorber was widely studied both experimentally~\cite{Genevet_prl08, Elsass_epjd10, Hachair_jstqe06, Hachair_pra09,Averlant_oe14, Averlant_sr16} and theoretically~\cite{VFK-JOB-99,FederovPRE2000, Bache2005, raey, Fedorov2007}.

Recently, much attention was paid to the investigation of the influence of delayed optical feedback on the stability properties of CSs. The impact of delayed feedback on the dynamics of CSs has been theoretically investigated in driven nonlinear optical resonators~\cite{TlidiPRL2009,PhysRevLett.110.014101,Tabbert_pre17} and broad-area VCSELs~\cite{Panajotov_epjd10,Tlidi_pra12,PimenovPRA13}. In particular, it was shown that nontrivial instabilities resulting in the formation of oscillons, CSs rings, labyrinth patterns, or moving CSs can develop. In addition, the influence of the feedback phase and carrier relaxation rate on the CS drift instability threshold was investigated. The influence of the optical delayed feedback on the dynamics of CSs in the generic Lugiato-Lefever model, describing the appearance of either spatial or temporal localized structures in nonlinear cavities was studied in~\cite{PanajotovPRA16}. A detailed investigation of the bifurcation structure of time-delayed feedback induced complex dynamics in a VCSEL with saturable absorption was performed recently in~\cite{Panajotov:14, PuzyrevPRA16}. It was demonstrated that the delayed optical feedback impacts the homogeneous lasing solution and the localized CS solutions in a similar way, causing oscillatory dynamics with either a period doubling or a quasiperiodic route to chaos, as well as a multistability of  stationary CSs. Furthermore, it was shown that at large delays apart from the drift and phase instabilities the soliton can exhibit a delay-induced modulational instability leading to a low-frequency switching of the CS motion.
  
In this paper we focus on the dynamical properties of CSs in the mean field model of a broad area VCSEL with saturable absorption and the impact of the delayed feedback on these properties. Using a combination of analytical, numerical, and path continuation methods we analyze the branches of stationary and moving CSs in the absence of the delayed feedback and discuss two types of traveling solutions, corresponding to  slowly and  fast moving CSs. We demonstrate that the fast moving CS can be interpreted as a lateral mode-locking regime in a broad-area laser with a single longitudinal mode and discuss how the delayed feedback impacts the dynamical behavior of the fast CS. We show that in the presence of the delay term, in addition to the drift and phase instabilities, CS can exhibit a delay-induced modulational instability associated with the translational neutral mode. A combination of the modulational and drift instabilities can lead to complex dynamics of the moving localized solution. Finally, we show that the delayed feedback can be effectively used to stabilize intrinsically moving CSs. 
\end{section}

\begin{section}{Model system}

 The dynamics of the slowly varying mean electric field envelope $E=E(r,\,t)$ and carrier densities $N=N(r,\,t)$,  $n=n(r,\,t)$, $r=(x,\,y)$ in active and passive part of the VCSEL subject to the delayed feedback can be described by the following system of equations~\cite{FederovPRE2000, Bache2005, raey}
 \begin{eqnarray}
\partial_t E &=& \left[(1-i\,\alpha)\,N+(1-i\,\beta)\,n-1+(d+i)\,\nabla_\perp^2\right]\,E+ \nonumber \\
&+&\eta\,e^{i\,\varphi}\,E(t-\tau) \label{eq:model1}\\
\label{eq:model2}
\partial_t N &=& b_1\left[\mu-N\,(1+|E|^2)\right] \\
\label{eq:model3}
\partial_t n &=& b_2\left[-\gamma-n\,(1+s\,|E|^2)\right]\,
\end{eqnarray}
Here, $\alpha\, (\beta)$ is the linewidth enhancement factor and $b_1\, (b_2 )$ is the ratio of photon lifetime to the carrier lifetime in the active (passive) layer, $\mu$ depends on the normalized injection current in the active material and $\gamma$ measures absorption in the passive material, $s$ is the saturation parameter and $d$ is the small diffusion (spatial filtering) coefficient. Time and transverse space coordinates are scaled to the photon life time and diffraction length, respectively. The optical feedback is modeled by single round trip delayed term characterized by the delay time $\tau$, feedback strength $\eta$ and phase $\varphi$. We assume that the external cavity is self-imaging, so that the diffraction of the feedback field can be neglected~\cite{Rosanov1975,LangKobayashi}. This means that the parameters $\eta$ and $\varphi$ are independent of the transverse coordinates. Note that the delay time $\tau$ is proportional to the external cavity length and is measured in units of the photon lifetime \cite{Panajotov:14}, whereas $\varphi$ describes a phase shift on the time scale of the fundamental lasing frequency, due to, e.g., shifting the mirror by a distance shorter than the wavelength of light.
\end{section}

\begin{section}{Bifurcation analysis of Cavity Solitons}\label{sec:BACS}
In the absence of the optical feedback term, i.e. for $\eta=0$,  the model~(\ref{eq:model1})-(\ref{eq:model3}) was extensively investigated in the literature~\cite{FederovPRE2000, Bache2005, raey,PhysRevA.84.053852,rohtua}. It possesses the trivial laser-off solution 
	\begin{equation}
		E = 0\,, \qquad N = \mu\,, \qquad n = -\gamma\,
	\end{equation}
which becomes unstable at the lasing threshold $\mu_\mathrm{th}=1+\gamma$.  In the bistability region $\mu_{\mathrm{fold}}<\mu<\mu_{\mathrm{th}}$, $\mu_{\mathrm{fold}}=\dfrac{(\sqrt{s-1}+\sqrt{\gamma})^2}{s}$, this solution co-exists with spatially homogeneous lasing solutions
\begin{equation}
		E = \sqrt{I}\,e^{i\omega t}, \quad N = \dfrac{\mu}{1+I}, \quad n = -\dfrac{\gamma}{1+s\,I},
	\end{equation}
corresponding to nonzero laser intensity $I=|E|^2$. In this region a non-trivial branch of CSs can be found~\cite{FederovPRE2000, Bache2005, raey}. While the shape of the CS branch is determined by the values of both linewidth enhancement factors $\alpha$ and $\beta$, the stability of CSs depends strongly on the ratio $b_1/b_2$ of the carrier lifetime in the active and in the passive medium~\cite{FederovPRE2000, Bache2005, rohtua}. Note that in the limit of instantaneous medium response, $b_{1,2}\rightarrow\infty$  and for $d=0$, linear operator describing the stability of the CS solution of  Eqs.~(\ref{eq:model1}-\ref{eq:model3}) has three zero eigenvalues, corresponding to the translational, phase and Galilean invariance. However, the Galilean transformation symmetry 
$$
E(r,\,t)\rightarrow E(r-vt,\, t) e^{i v x/2-i v^2 t / 4}
$$
is typically broken for finite relaxation rates leading to a shift of the corresponding real eigenvalue from the origin in the complex plane. As a result, for small non-vanishing values of $b_{1,2}^{-1}$, a CS can loose the stability with respect to a drift-bifurcation, giving rise to a CS moving with a constant velocity $v$. We refer to these structures as slow moving CSs (SCS). Note that the non-vanishing diffusion coefficient $d$ also breaks the Galilean transformation symmetry leading to the shift of the corresponding eigenvalue from the origin. In this case, however, it is shifted to the negative half plane of the real axis, so that the linear stability of CSs is not affected by $d$. When $b_1$, $b_2$ are small enough, different bifurcation scenario leading to various types of Andronov-Hopf bifurcations and their interactions can be observed. Furthermore, numerical simulations performed in ~\cite{FederovPRE2000,Fedorov2007} revealed the existence of CSs moving with large velocities, the so-called fast CSs (FCS). These structures can coexist with either stationary or slow moving CSs and are characterized by narrower intensity distribution and much greater peak intensities and velocities than those of SCSs.  

\paragraph*{Bifurcation analysis in one dimension} In this subsection, we perform detailed bifurcation analysis of both SCSs and FCSs solutions in one spatial dimension. First, we consider the formation of stable stationary CSs. One can find them e.g., numerically by direct numerical integration of (\ref{eq:model1})-(\ref{eq:model3}) with some fixed value of $\mu$. The result of the numerical simulations obtained after sufficiently long integration interval can be then used as an initial guess of the Newton method for the solution of nonlinear stationary problem (\ref{eq:model1})-(\ref{eq:model3}) with time derivatives set to zero and spatial derivatives discretized according to some finite difference scheme. Since all the moving CSs in the system (\ref{eq:model1})-(\ref{eq:model3}) have constant velocity $v$, we can use almost the same Newton's algorithm for continuation of their branches by looking for the one-dimensional CS in the form
\begin{equation}\label{eq:movsol}
E(x,\,t) = E(\xi) e^{i \omega t},
\end{equation}
where $\xi = x - v t$, $v$ is the unknown speed and $\omega$ is the unknown frequency shift. We note, however, that for the FCS with $v \gg 1$ direct numerical simulation of (\ref{eq:model1})-(\ref{eq:model3}) does not provide suitable initial guess for the Newton method independently on the simulation time. Indeed, when the symmetry with respect to the Galilean transformation is only slightly broken the phase of the FCS moving at high velocity $v \gg 1$ changes very rapidly in space and time, which would require the use of very small discretization step in the Newton's algorithm. To eliminate this fast change of the phase, we estimate the velocity of the FCS with the help of direct numerical simulation , $v\approx \tilde v$ and perform the transformation of the field
$E(x, t) := E(x, t) e^{i x \tilde v / 2 - i t \tilde v^2/ 4}$. Then we obtain the following system for the determination of the amplitude of the FCS:
\begin{eqnarray}
 0&=&  (i + d) \Delta E - (v - i d \tilde v) \frac{\partial E}{\partial \xi}  -(1+i\omega)\,E \notag \\ && + \left((1 - i \alpha) N + (1 - i \beta) n  - (i + d)  \frac{\tilde v^2}{4} \right) E,  \label{eq:en} \\
0&=& -v \frac{\partial N}{\partial \xi} + b_1(\mu - (1 + |E|^2) N),\label{eq:gn}\\
0&=&  - v \frac{\partial n}{\partial \xi} +b_2( -\gamma - (1 + s |E|^2) n).\label{eq:an}
\end{eqnarray}
We have solved Eqs.~(\ref{eq:en})-(\ref{eq:an}) using the Newton method with the additional auxiliary condition $E(\tilde x) = \tilde E$, where $\tilde x$ is a point at half-maximum of the CS, and $\tilde E$ is the field obtained numerically at this point. In this way, branches of both SCS and FCS can be reconstructed and the resulting bifurcation diagram is presented in Fig.~\ref{fig:bdss}~(a) with a closeup in panel (b).
\begin{figure}[!ht]
\begin{center}
\begin{tabular}{cc}
\includegraphics*[width=0.24\textwidth]{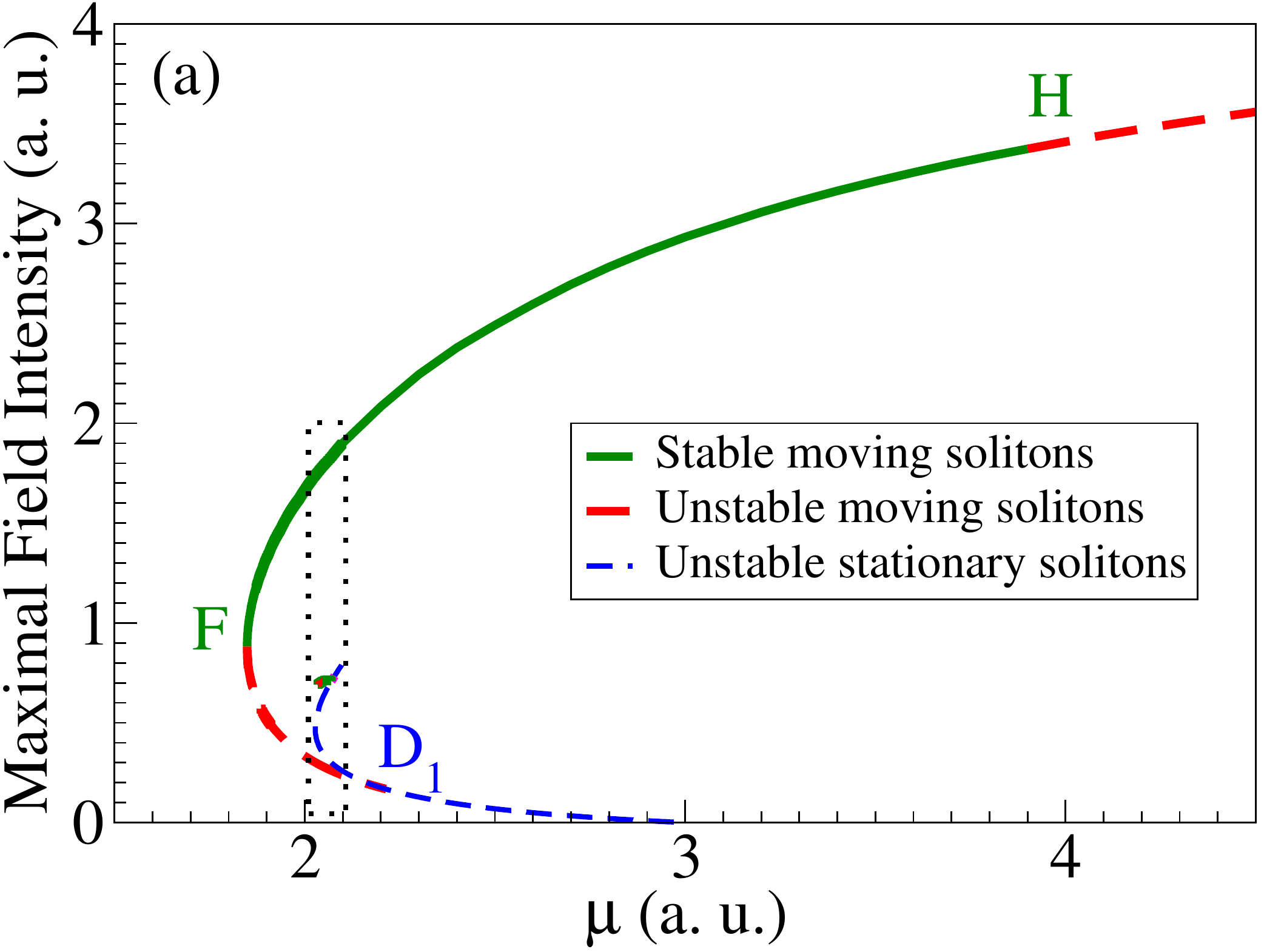}&
\includegraphics*[width=0.24\textwidth]{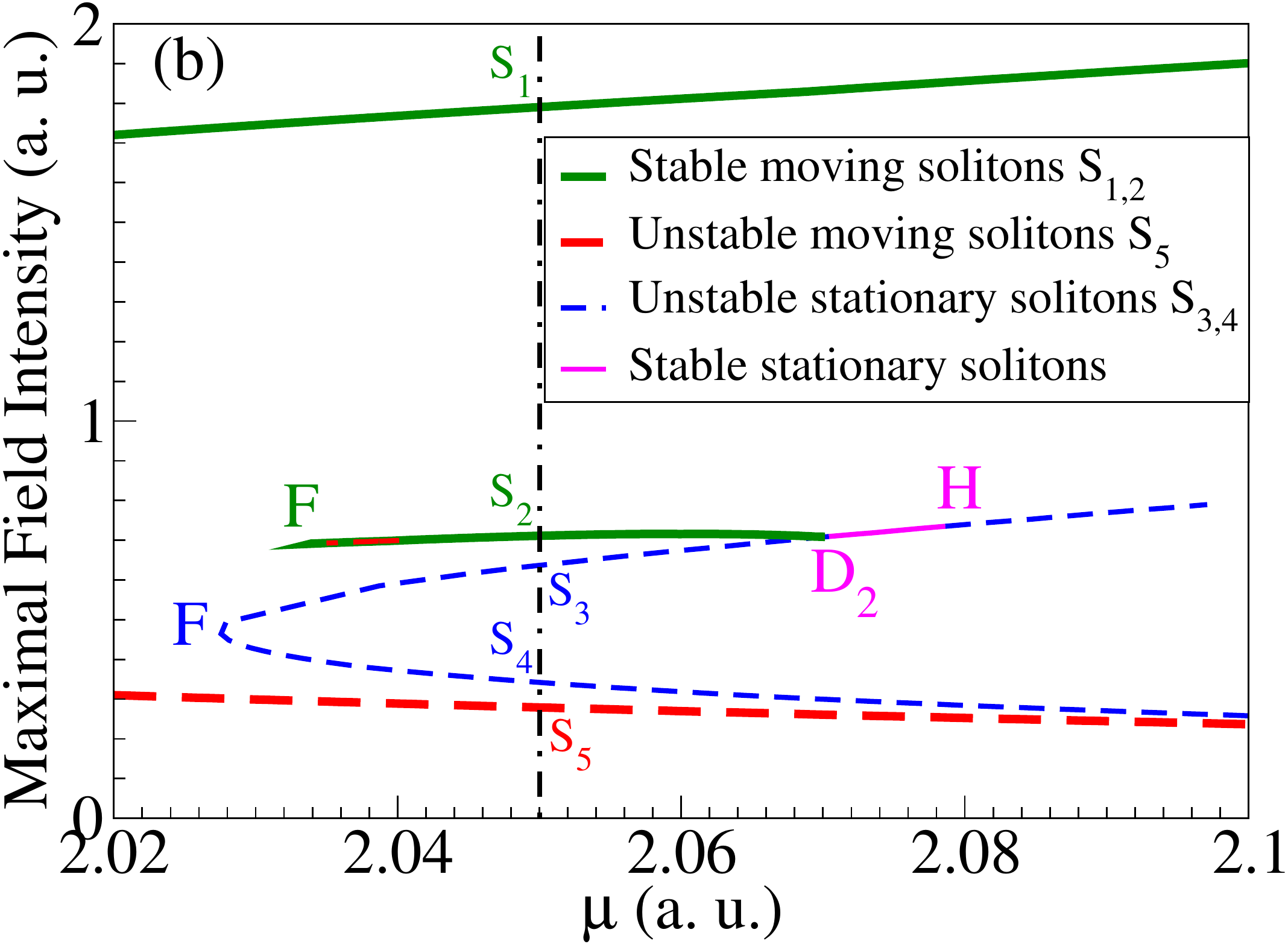}\\[2mm]
\includegraphics*[width=0.24\textwidth]{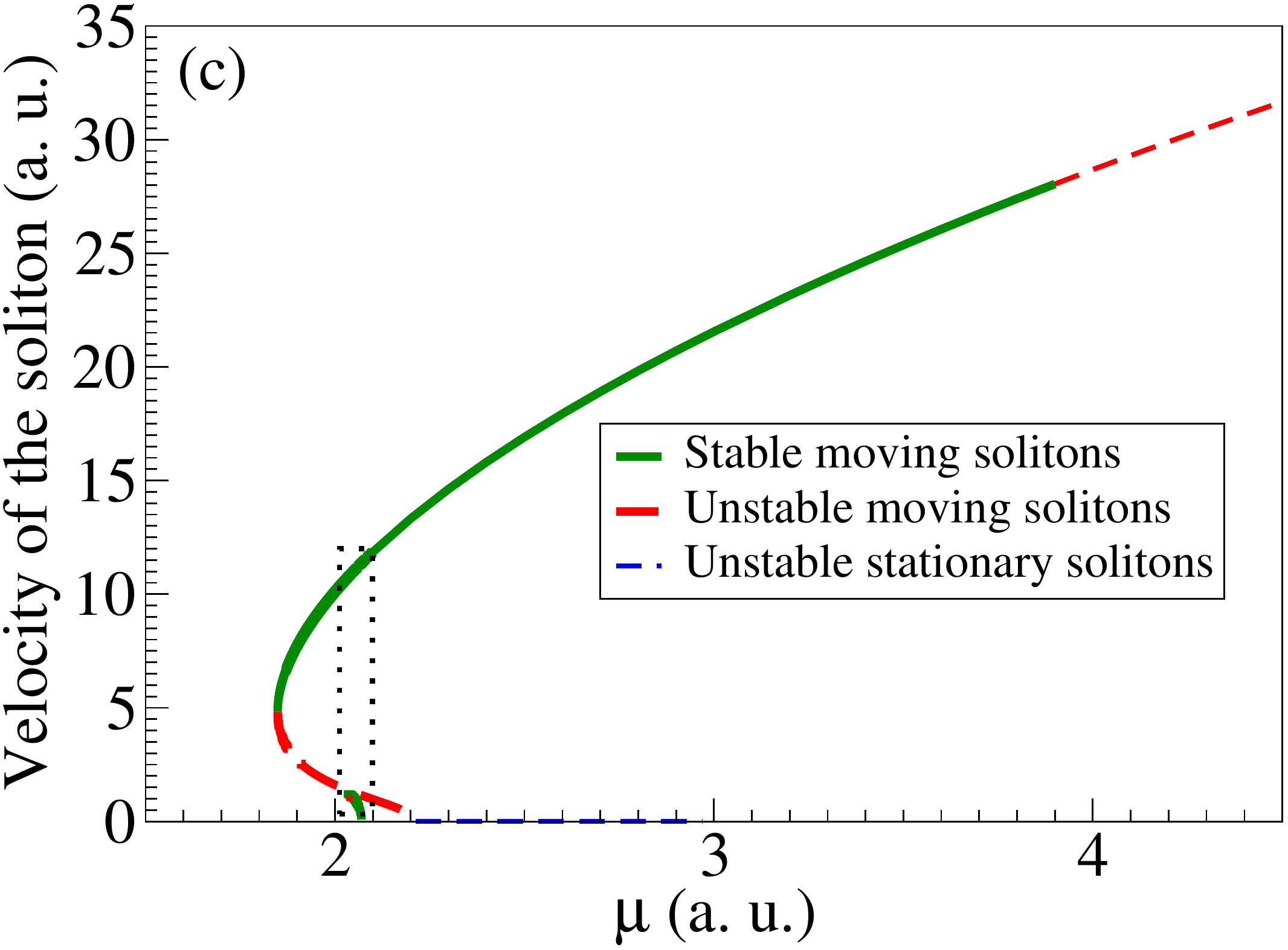}&
\includegraphics*[width=0.24\textwidth]{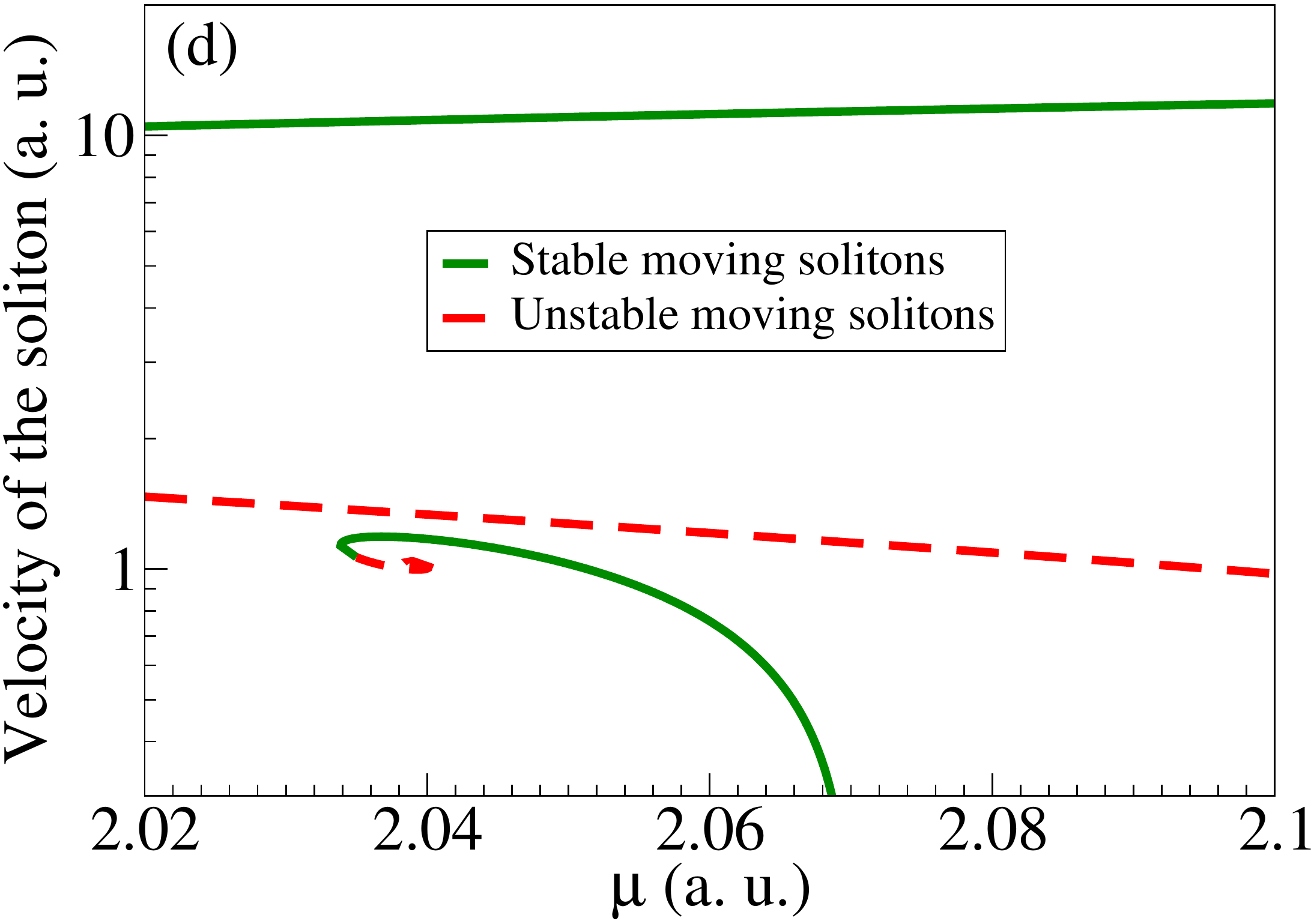}
\end{tabular}
\end{center}
\caption{One-dimensional bifurcation diagram for a single CS (top, a-b) and its velocity on the corresponding branches (bottom, c-d). Here, $b_1 = 1.0, b_2 = 3.33, d = 0.01$, $\alpha=\beta = 0, \gamma = 2,  s = 10$ and $\mu$ is varied in the full region of existence of stable CSs at the left (a,c) and in the zoomed region (denoted by the dotted box at the insets (a,c) ) of stable stationary and SCSs at the right (b,d). $\mathrm{F}$, $\mathrm{D}_{1,2}$, and $\mathrm{H}$ denote fold, drift and Andronov-Hopf bifurcations, correspondingly, bold solid lines correspond to stable moving CSs (top - to fast CSs, bottom - to slow CSs), thin solid line  between points $\mathrm{D}_2$ and $\mathrm{H}$ on (b) - to stable stationary CSs, and dashed lines correspond to unstable CSs. Dash-dotted line in the inset (b) denotes the value $\mu=2.05$ where five coexisting solitons S$_1$-S$_5$ can be observed.}
\label{fig:bdss}
\end{figure}
 One can see from Fig.~\ref{fig:bdss}~(b) that the stability range of the stationary CS is limited by two bifurcation points, Andronov-Hopf point (H) at $\mu \approx 2.08$ and pitchfork drift bifurcation point $(\mathrm{D}_2)$ at $\mu \approx 2.07$. One can continue the branch of an unstable stationary CS by decreasing the parameter $\mu$ below the drift bifurcation (see Fig. \ref{fig:bdss} (b)) to observe that the branch of an unstable CS bifurcates from the laser-off state with zero laser intensity at the lasing threshold  $\mu =\mu_{\mathrm{th}}= 3$ (see Fig. \ref{fig:bdss} (a)). Further, by continuation of the branch of the SCS near the drift bifurcation (see Fig. \ref{fig:bdss} (b))  one can see that they exist only in a small range of the pump rates $2.03 < \mu < 2.07$, and their velocity is an order of magnitude smaller than that of the FCS in the same parameter range (see Fig. \ref{fig:bdss} (d), below the dashed line). 

In addition, one can explore the origin of the branch of FCSs by following this branch from the same region near $\mu = 2.06$ (see Fig. \ref{fig:bdss} (a), (c)) towards lower values of $\mu$. One can  see that the FCS looses stability at a saddle-node bifurcation at $\mu \approx 1.9$. After this bifurcation the unstable FCS branch can be continued until it merges with the branch of the unstable stationary CS at a drift bifurcation point $\mathrm{D}_1$ at $\mu \approx 2.2$. Therefore, our continuation results indicate that the branch of the FCS originates from the unstable part of the stationary branch of the CS. Furthermore, by increasing $\mu$ we see that FCSs can exist far above the lasing threshold $\mu_{\mathrm{th}} = 3$, and their stability is lost at an Andronov-Hopf bifurcation, where oscillating fast solutions can be observed~\cite{FederovPRE2000}. 
 \begin{figure}[!ht]
\begin{center}
\begin{tabular}{cc}
\includegraphics*[width=0.24\textwidth]{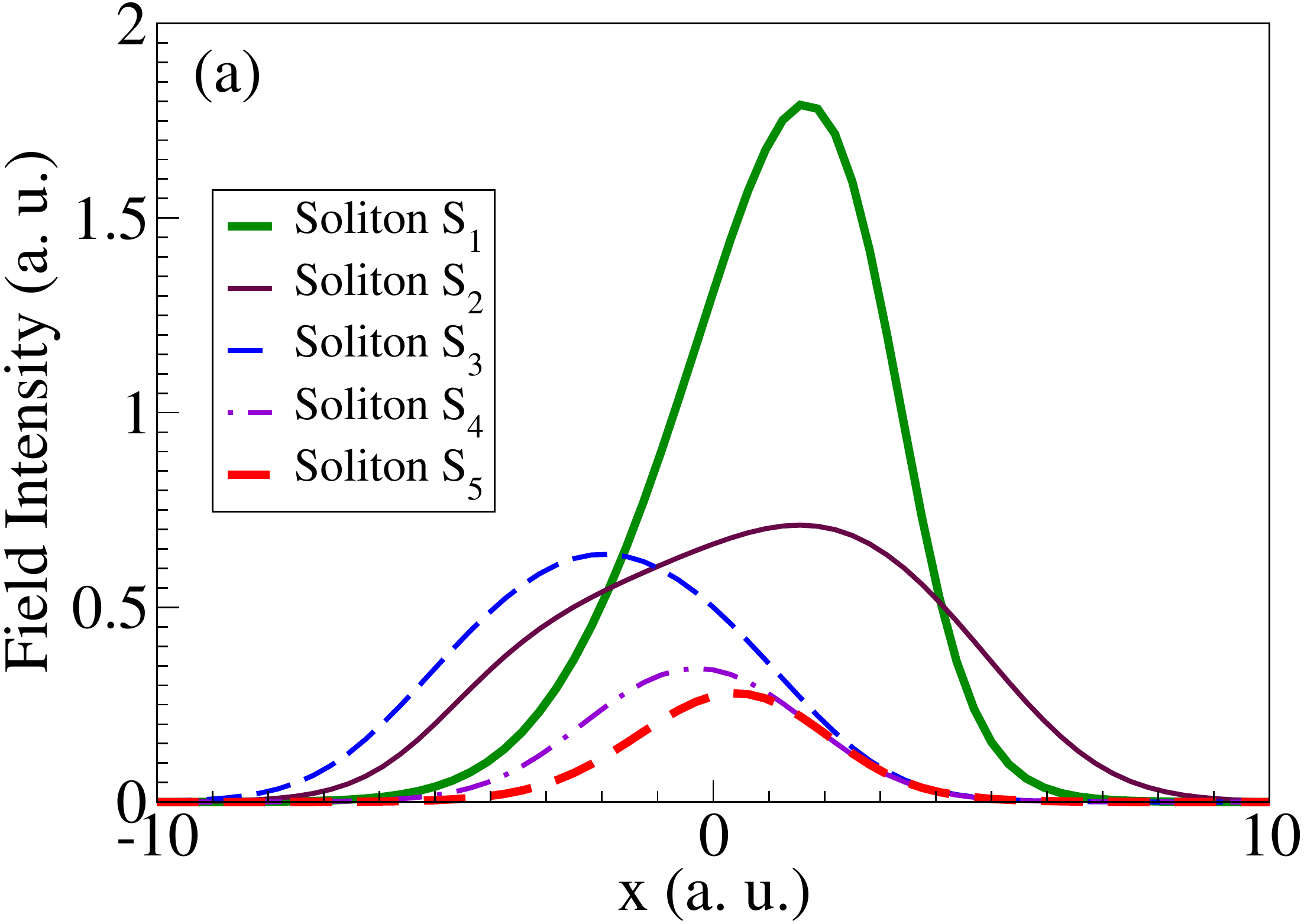}&
\includegraphics*[width=0.24\textwidth]{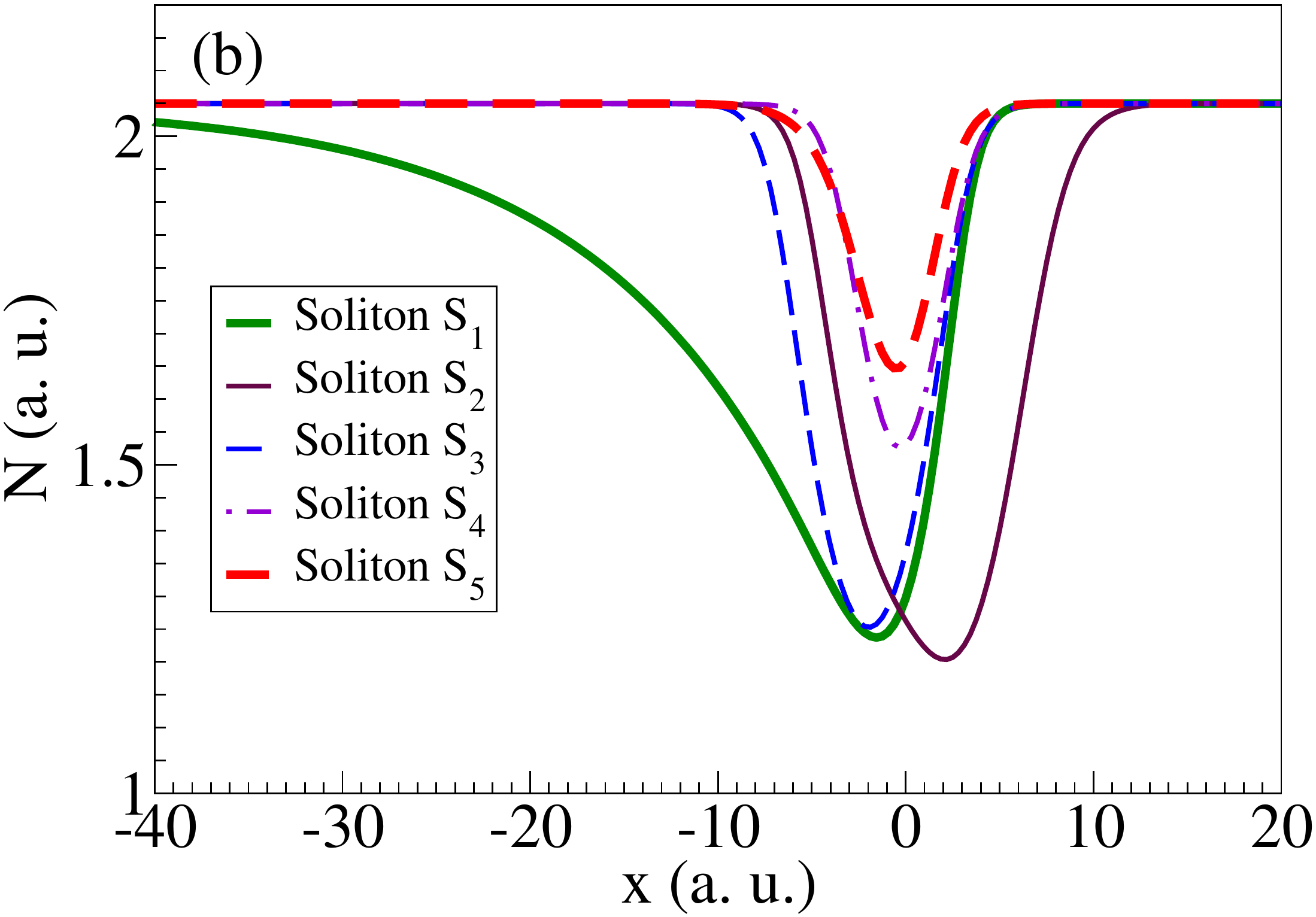}\\[2mm]
\includegraphics*[width=0.24\textwidth]{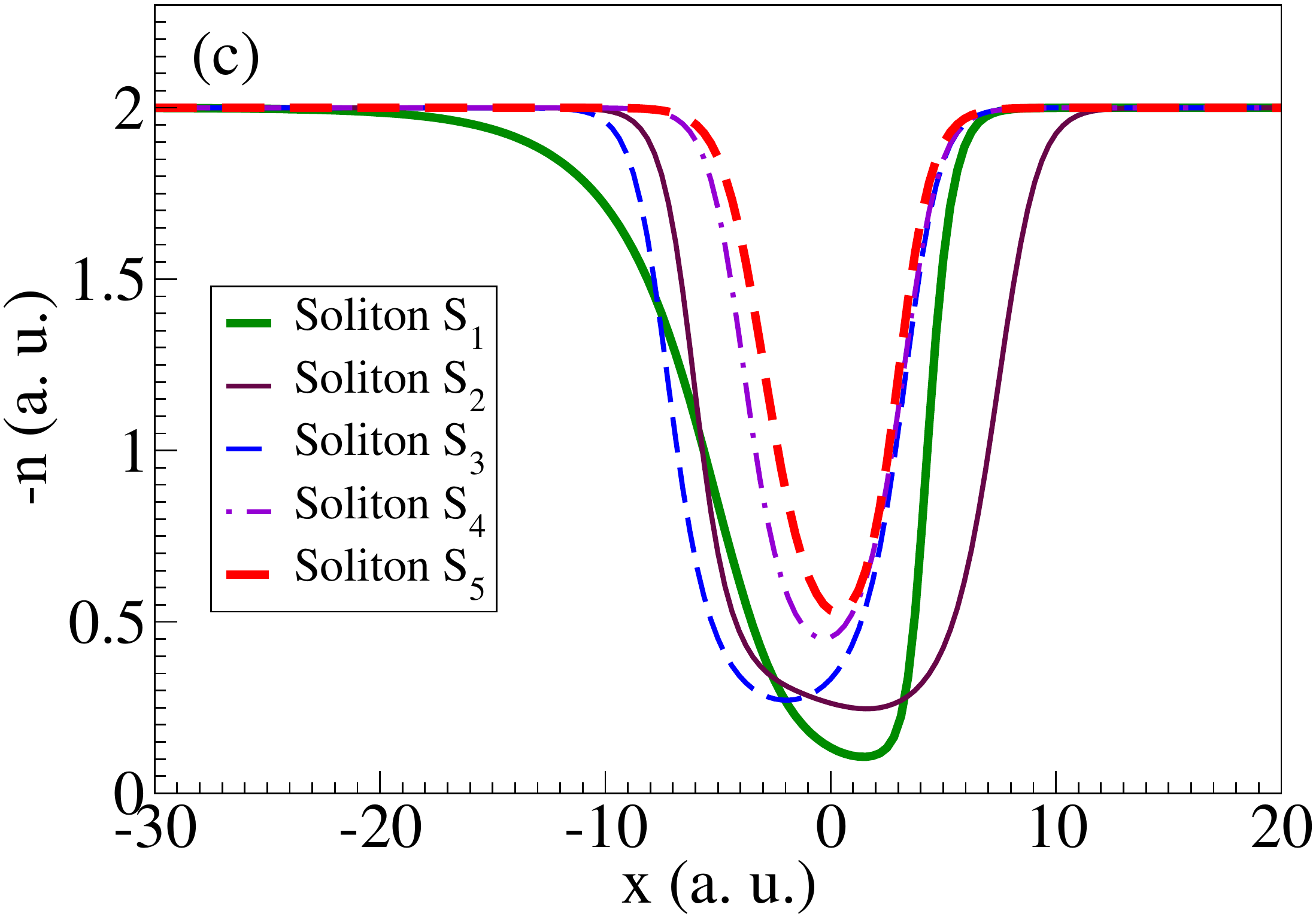}&
\includegraphics*[width=0.24\textwidth]{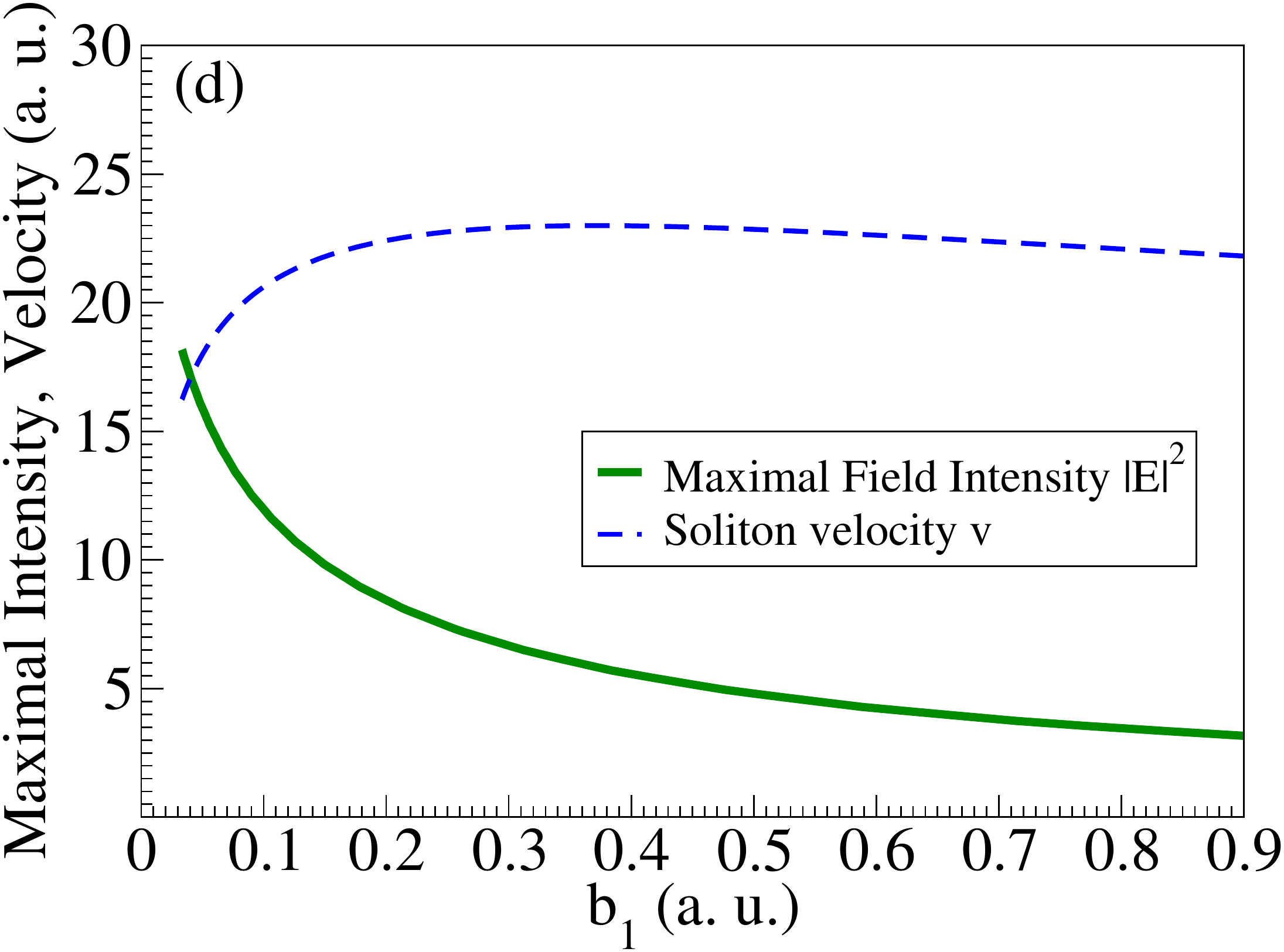}
\end{tabular}
\end{center}
\caption{(a) A field intensity $|E(x)|^2$, (b) carrier densities $N(x)$ (c) $-n(x)$ for the solitons S$_1$-S$_5$ marked at the bifurcation diagram on Fig.~\ref{fig:bdss} (b). (d) Maximal field intensity and velocity of the FCS S$_1$ for various values of $b_1$ at $\mu_{\mathrm{th}} = 3$. Other parameters are the same as in Fig.~\ref{fig:bdss}.}
\label{fig:sols}
\end{figure}
 Transverse intensity distributions of stable and unstable solitons calculated for the same pump parameter value $\mu = 2.05$  (cf. Fig. \ref{fig:bdss} (b)) and corresponding spatial profiles of the gain and absorption are shown in Fig. \ref{fig:sols} (a)-(c). Note that the shape of both unstable stationary CSs S$_3$, S$_4$ shown in Fig. \ref{fig:sols} is symmetric. On the other hand, stable and unstable fast moving CSs S$_1$ and S$_2$ demonstrate a strong asymmetry of the soliton profile, which reflects the movement direction. Furthermore, the stable FCS S$_1$ exhibits rather slow gain and absorption recovery on the trailing tail of the soliton typical for the mode-locking regime. Stable localized structures that stem from fundamental (or harmonic) mode-locking regime and coexist with a stable laser off-state below the lasing threshold were recently reported in~\cite{MJB-PRL-14}. Similarly to the situation presented here, the mode-locking regime corresponding to these structures becomes self-starting above the lasing threshold. Finally, we decrease the ratio $b_1/b_2$ of the gain and the absorber relaxation rates to the value typical for mode-locked semiconductor lasers~\cite{VT-PRA-05, pimenov2014effect}, and observe that FCSs are preserved (see Fig.~\ref{fig:sols}~(d)). Therefore, we conclude that stable fast solitons, which can exist both below and above the lasing threshold, might be interpreted as \emph{a lateral mode-locking regime} in a broad-area laser with a single longitudinal mode. 
 
\paragraph*{FCSs in two dimensions} In this subsection, we briefly discuss the results of numerical calculations using the path-following algorithm of  two-dimensional FCSs that were first predicted in~\cite{Fedorov2007} in the limiting case $b_2 = \infty$. We use the soliton calculated by direct numerical integration of Eqs.~\eqref{eq:model1}-\eqref{eq:model3} in two spatial dimensions as an initial guess and perform the same continuation steps as in the one-dimensional case. In this way a branch of  FCSs can be found and continued till the lasing threshold $\mu = \mu_{\mathrm{th}}$, as it is shown in Fig.~\ref{fig:bif2D}~(a). It can be seen from this figure that this branch exists only below the threshold $\mu_{\mathrm{th}}$.
\begin{figure}[!ht]
\begin{center}
\begin{tabular}{cc}
\includegraphics*[width=0.24\textwidth]{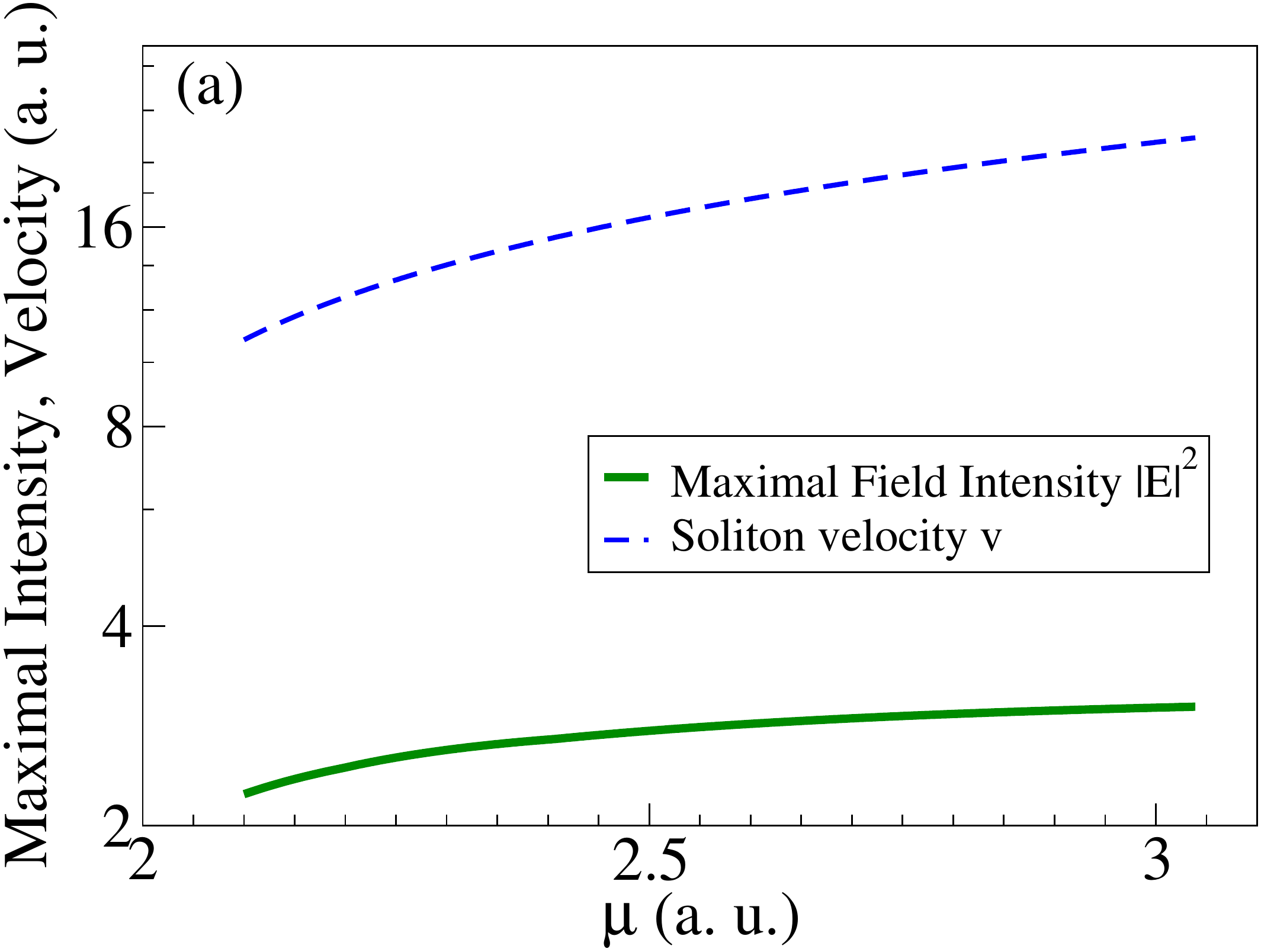}&
\includegraphics*[width=0.24\textwidth]{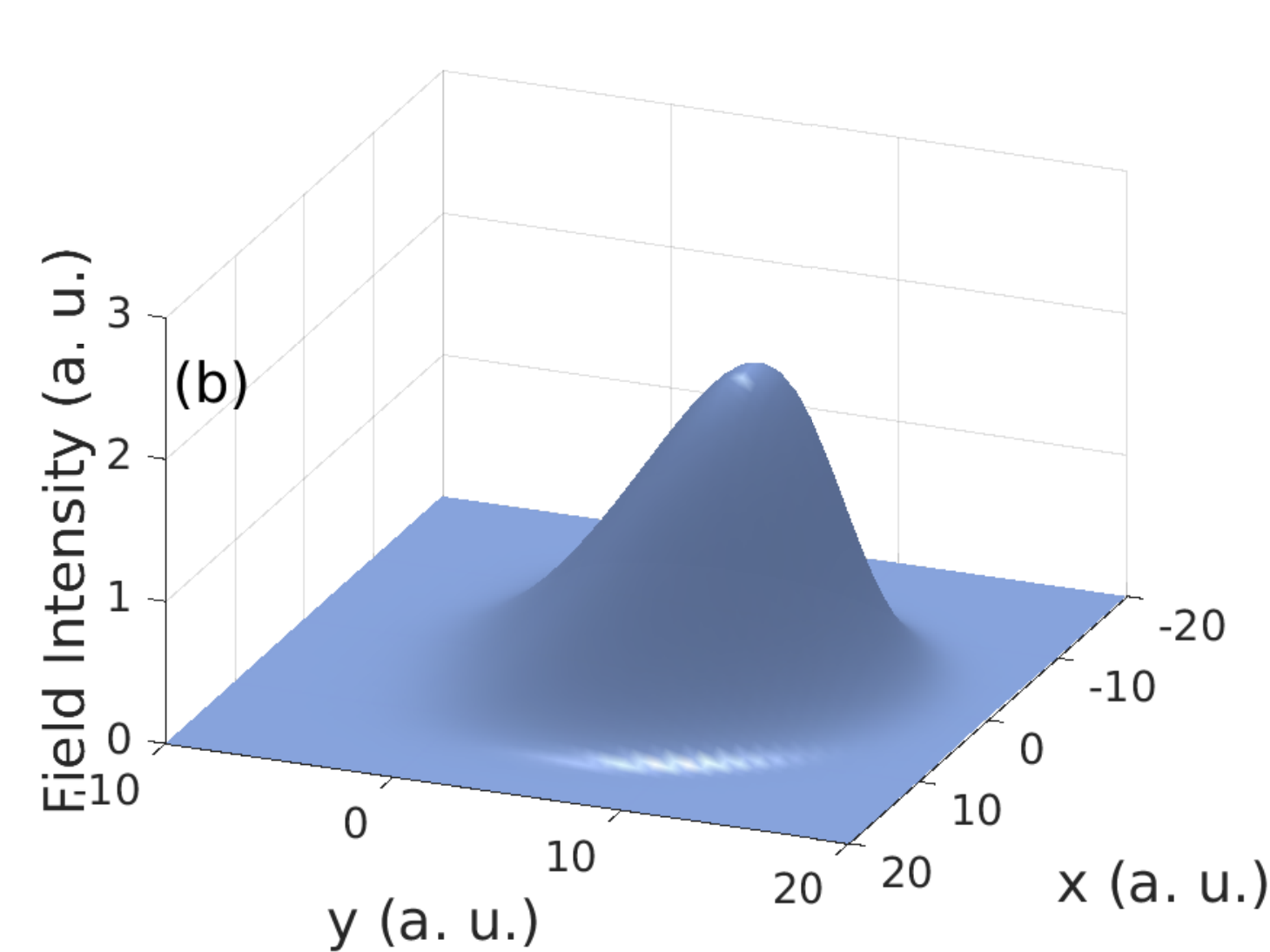}\\[2mm]
\includegraphics*[width=0.24\textwidth]{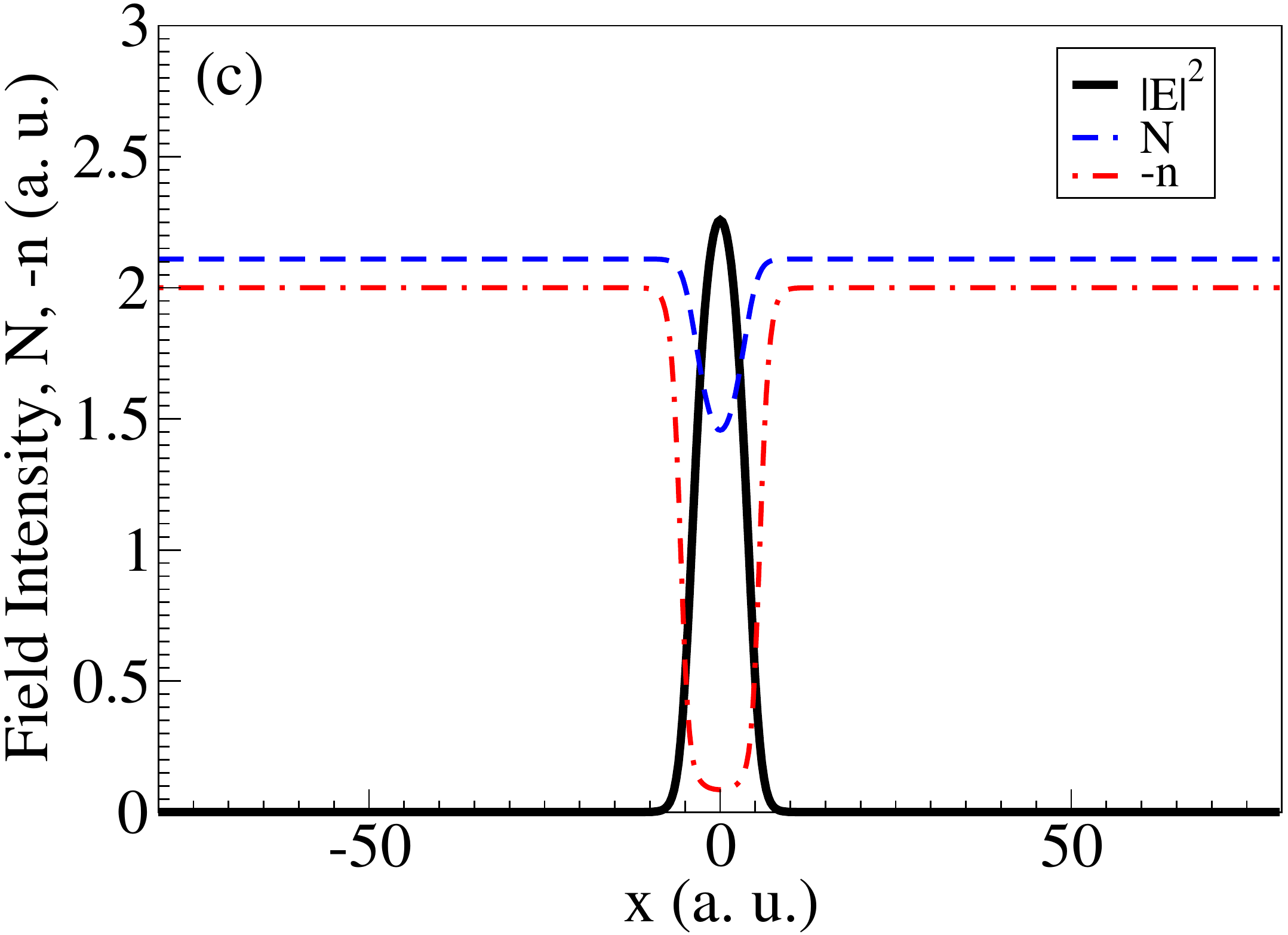}&
\includegraphics*[width=0.24\textwidth]{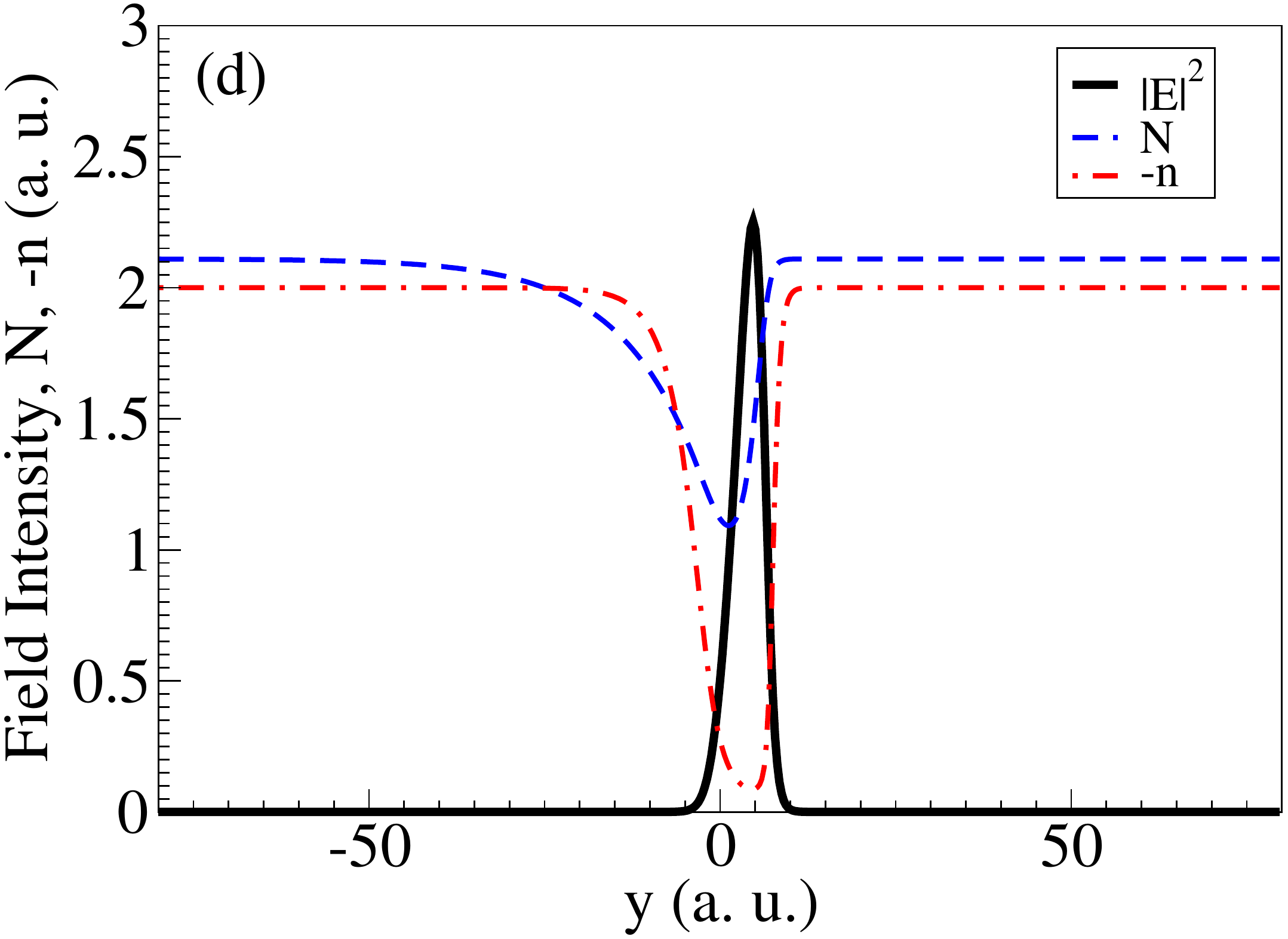}
\end{tabular}
\end{center}
\caption{(a) Peak intensity (solid) and velocity (dashed) corresponding to a branch of a two-dimensional FCS as functions of the the pump rate $\mu$ below the lasing threshold $\mu_{\mathrm{th}}=3$. (b) Intensity profile of a two-dimensional FCS calculated at $\mu = 2.11$. Panels (c) and (d) show typical cross-section profiles of the FCS in $x$- and $y$-directions, respectively. Other parameter are the same as in Fig. \ref{fig:bdss}.}
\label{fig:bif2D}
\end{figure}
Using direct numerical integration of the system \eqref{eq:model1}-\eqref{eq:model3} we found that for $\mu = 2.11$ two-dimensional FCS is stable (see Fig. \ref{fig:bif2D}~(b)-(d)). However at slightly larger $\mu$ it becomes unstable with respect to an Andronov-Hopf bifurcation. Therefore, the most part of the  branch shown in Fig.~\ref{fig:bif2D}~(a) corresponds to unstable FCSs. As one can see from Fig. \ref{fig:bif2D}~(c), the FCS profile remains symmetric in the $x$-direction similarly to the stationary one-dimensional CS and it is strongly asymmetric along the direction of the soliton motion ($y$). In this direction the FCS demonstrates the same mode-locking behavior as in the one-dimensional case discussed above (see Fig. \ref{fig:bif2D}~(d)). We note that all the properties of the two-dimensional FCS strongly resemble the behavior of two-dimensional light bullets observed in passively mode-locked semiconductor lasers~\cite{J-PRL-16} with the difference, however, that for the FCS shown in Fig.~\ref{fig:bif2D} the mode-locking is associated with the \emph{transverse direction} $y$ instead of longitudinal direction.  
\end{section}

\begin{section}{Delay-Induced Drift of Cavity Solitons}\label{sec:DelIndDyn}

\paragraph*{Solution structure}

 Now let us switch to the case of nonzero feedback strength. In order to find localized solutions of Eqs.~ (\ref{eq:model1})-(\ref{eq:model3}) for $\eta\neq 0$, we substitute the ansatz 
 $E=E_0(x)\,e^{-i\,\omega\,t}$, $N=N_0(x)$, $n=n_0(x)$ into Eqs.~ (\ref{eq:model1})-(\ref{eq:model3}) and obtain the following  set of ordinary differential equations for   
unknowns $E_0,\, N_0,\, n_0$ and $\omega$
 \begin{eqnarray}
 \nonumber
  0&=&\left((1-i\,\alpha)\,N_0+(1-i\,\beta)\,n_0-1+i\,\omega\right)\,E_0\\
   &+& (d+i)\,\nabla^2+\eta\,e^{i\,\theta}\,E_0\,, \label{eq:StatSol1} \\ 
   0&=&b_1\left(\mu-N_0\,(1+|E_0|^2)\right)\,,\label{eq:StatSol2}\\
  0&=&-b_2\left(\gamma+n_0\,(1+s\,|E_0|^2)\right)\,. \label{eq:StatSol3}
 \end{eqnarray}
where $E_0$ is the complex amplitude with the field intensity $|E_0|^2$, $\omega$ is the frequency shift, whereas
\begin{equation}\label{eq:effphase}
 \theta=\omega\,\tau+\varphi\,\, \mathrm{mod}\, 2\pi
\end{equation}
denotes the effective feedback phase~\cite{PuzyrevPRA16}. The system~\eqref{eq:StatSol1}-\eqref{eq:StatSol3} defines a set of CS solutions parametrized by the phase $\theta$. Since Eqs.~\eqref{eq:StatSol1}-\eqref{eq:StatSol3} are ordinary differential equations, this set can be found with the help of numerical continuation techniques in a similar way as in the system without delay. 
This yields the sets of CSs corresponding to different fixed values of the phase $\theta$ of the delayed field. Taken all together they form a tube shaped manifold of all possible localized solutions of the system with a given delay strength~\cite{PuzyrevPRA16}. Then, for any given $\tau$ and $\varphi$ one can determine the actual branches by solving equation (\ref{eq:effphase}) implicitly on this manifold. One can observe that this branch performs a number of turns around the tube giving rise to multistability of CSs, so that at the fixed value of the pump parameter $\mu$ and sufficiently large delay time $\tau$, one obtains a discrete set of external cavity solitons, similarly to the case of external cavity modes in a single-mode laser with delayed feedback~\cite{GreenSIAM2009,YaWolSIAM2010,SorianoRMoPhys2010,PuzyrevSIAM2014}. Note that a similar multistability effect was experimentally observed in a broad-area VCSEL with frequency-selective feedback ~\cite{TanguyPRA2006}. The existence of such a multistability follows explicitly from the form of Eq. (\ref{eq:effphase}). In particular, in~\cite{PuzyrevPRA16} it was shown that the number of solutions grows linearly with the delay time, so that in the limit of large delay time, the solutions cover the whole range of the effective phases $\theta$ with the distance between them of the order $1/\tau$ ~\cite{YaWolSIAM2010,PuzyrevSIAM2014}.

\paragraph*{Linear stability analysis}
In order to analyze the stability of a CS solution in the presence of delayed feedback, we linearize the system~\eqref{eq:StatSol1}-\eqref{eq:StatSol3} around $\mathbf{q}_0(x)=(\mathrm{Re}(E_0),\,\mathrm{Im}(E_0),\,N_0,\,n_0)^T$ and arrive at the transcendental eigenvalue problem:
\begin{equation}
\label{eq:EigenProblem}
 \left(\mathfrak{L}'(\mathbf{q}_0)-\lambda\,\mathrm{I}+\eta\mathrm{B}\,e^{-\lambda\,\tau}\right)\,\boldsymbol{\psi}=0\,,
\end{equation}
where $\boldsymbol{\psi}$ is an eigenfunction, corresponding to the eigenvalue $\lambda$, $\mathfrak{L}'(\mathbf{q}_0)$ is the linearization operator,  $\mathrm{B}$ is a rotation matrix by the phase angle $\theta$ and $\mathrm{I}$ is the identity matrix. 

As we mentioned above, the linearization operator $\mathfrak{L}'$ possesses two zero eigenvalues corresponding to the even phase shift neutral eigenfunction (Goldstone mode) $\boldsymbol{\psi}_{\mathrm{ph}}=i\,\mathbf{q}_0$ as well as to the odd Goldstone mode $\boldsymbol{\psi}_{\mathrm{tr}}=\partial_{\mathbf{x}}\mathbf{q}_0$.  Note that the Galilean transformation symmetry is broken due to the presence of diffusion $d$, delayed feedback term, and non-instantaneous medium response~\cite{VFK-JOB-99,FederovPRE2000, PuzyrevPRA16}. 
 
Either drift or phase bifurcation can occur when the critical real eigenvalue passes through zero at the bifurcation point, so that the corresponding critical eigenfunction at this point is proportional to one of the two neutral eigenfunctions $\boldsymbol{\psi}_0=\boldsymbol{\psi}_{\mathrm{tr},\mathrm{ph}}$. This critical eigenvalue can be either a delay-induced branch of the corresponding neutral eigenvalue or, in the case of intrinsic drift bifurcation (i.e., drift bifurcation in the absence of delayed feedback), correspond to a Galilean mode.

Since both drift and phase bifurcations occur when the eigenvalue corresponding to the corresponding neutral mode becomes doubly degenerate with geometric multiplicity one, in the vicinity of the bifurcation point we have $\lambda=0+\varepsilon,\quad \boldsymbol{\psi}=\boldsymbol{\psi}_0+\varepsilon\,\boldsymbol{\psi}_1, \quad \varepsilon\ll 1$ with some unknown function $\boldsymbol{\psi}_1$. Substituting this ansatz into Eq.~\eqref{eq:EigenProblem}, expanding the resulting equation into power series in $\varepsilon$, and collecting first order terms one obtains 
$$
\left(\mathfrak{L}'(\mathbf{q}_0)+\eta\mathrm{B}\right)\,\boldsymbol{\psi}_1=\left(\mathrm{I}+\eta\,\tau\,\mathrm{B}\right)\boldsymbol{\psi}_0\,.
$$ 
The existence of a non-trivial solution of this equation is equivalent to the following solvability condition leading to the general expression for the onset of drift and phase bifurcations~\cite{PuzyrevPRA16} :
 \begin{equation}\label{eq:drift_phase_bif}
   \eta\tau=-\frac{<\boldsymbol{\psi}^{\dag}_0|\boldsymbol{\psi}_0>}{<\boldsymbol{\psi}^{\dag}_0|\mathrm{B}\,\boldsymbol{\psi}_0>}\,.
 \end{equation}
Note that in the case when $\mathrm{B}=\mathrm{I}$, Eq.~\eqref{eq:drift_phase_bif} reduces to the simple condition $\eta\tau=-1$ first obtained for the Swift-Hohenberg equation with delayed feedback term~\cite{TlidiPRL2009,PhysRevLett.110.014101}.

 \paragraph*{Drift Bifurcation}
 The components of the real four-dimensional vector-function $\mathbf{q}_0$ representing a stationary localized solution of the system~(\ref{eq:model1})-(\ref{eq:model3}) obey the following relations
 $$
q_3=\frac{\mu}{1+q_1^2+q_2^2}\,,\qquad q_4=-\frac{\gamma}{1+s\,(q_1^2+q_2^2)}\,,
$$
where we define $\mathbf {q_0}=(q_1,\,q_2,\,q_3,\,q_4)^T=(\mathrm{Re}(E_0),\,\mathrm{Im}(E_0),\,N_0,\,n_0)^T$. Furthermore, one can show that 
\begin{eqnarray}
 \psi_3&=&-\frac{2\,\mu\,(q_1\,\psi_1+q_2\,\psi_2)}{(1+q_1^2+q_2^2)^2}\label{3}\,,\\
 \psi_4&=&\frac{2\,s\,\gamma\,(q_1\,\psi_1+q_2\,\psi_2)}{\left(1+s(q_1^2+q_2^2)\right)^2}\,,
\end{eqnarray}
and 
\begin{eqnarray}
 \psi^{\dag}_3=\frac{q_1\,\psi^{\dag}_1+q_2\,\psi^{\dag}_2+\alpha\,(\psi^{\dag}_1\,q_2-\psi^{\dag}_2\,q_1)}{b_1\,(1+q_1^2+q_2^2)}\,,\\
 \psi^{\dag}_4=\frac{-q_1\,\psi^{\dag}_1-q_2\,\psi^{\dag}_2+\beta\,(\psi^{\dag}_2\,q_1-\psi^{\dag}_1\,q_2)}{b_2\,(1+s\,(q_1^2+q_2^2))}\,,\label{4c}
\end{eqnarray}
where $\boldsymbol{\psi}_0=(\psi_1,\,\psi_2,\,\psi_3,\,\psi_4)^T$ is the neutral translational eigenfunction and $\boldsymbol{\psi}^{\dag}_0=(\psi^{\dag}_1,\,\psi^{\dag}_2,\,\psi^{\dag}_3,\,\psi^{\dag}_4)^T$ is the corresponding adjoint neutral eigenmode. Hence, the threshold~\eqref{eq:drift_phase_bif} of the drift bifurcation can be written in terms of the first two components of the vectors  $\boldsymbol{\psi}_0,\,\boldsymbol{\psi}_0^{\dag}$:
\begin{equation}
  \eta\tau=-\frac{\sum\limits_{j=1}^4<\psi^{\dag}_j|\psi_j>}{p_1\,\cos\theta+p_2\,\sin\theta}\,.
\end{equation}
 where $\psi_{3,4}$ and $\psi_{3,4}^{\dag}$ are defined by Eqs.~(\ref{3})-(\ref{4c}) and
 $$
 p_1=<\psi^{\dag}_1|\psi_1>+<\psi^{\dag}_2|\psi_2>,\, p_2=<\psi^{\dag}_2|\psi_1>-<\psi^{\dag}_1|\psi_2>.
 $$
 
 \paragraph*{Phase Bifurcation and Multistability}
 Similar to the translational zero eigenvalue, the second zero eigenvalue corresponding to the phase-shift symmetry can become doubly degenerate in the presence of delayed feedback. While in the case of the drift bifurcation, a drift-pitchfork bifurcation takes place, the phase bifurcation corresponds to a saddle-node bifurcation, where a pair of CS solutions merge and disappear. Note that this fold condition follows directly from Eq.~\eqref{eq:effphase} and can be written as~\cite{PuzyrevPRA16} 
	\begin{equation} \label{eq:dwdt}
		\dfrac{d\omega}{d\theta}=\dfrac{1}{\tau}\,.
	\end{equation}
Indeed, a differentiation of Eqs.~\eqref{eq:StatSol1}-\eqref{eq:StatSol3} with respect to $\theta$ yields
$$
\left(\mathfrak{L}'(\mathbf{q}_0)+i\omega+\eta e^{i\theta}\right)\frac{\partial\mathbf{q}_0}{\partial\theta}=-\frac{d\omega}{d\theta}\boldsymbol{\psi}_{\mathrm{ph}}-\eta\,\mathrm{B}\boldsymbol{\psi}_{\mathrm{ph}}\,.
$$
The solvability condition for this equation, which determines the onset of CS phase bifurcation, is equivalent to the expression~\eqref{eq:drift_phase_bif} with $\boldsymbol{\psi}_0=\boldsymbol{\psi}_{\mathrm{ph}}$
\begin{equation}
 \eta=-\frac{<\boldsymbol{\psi}^{\dag}_{\mathrm{ph}}|\dfrac{d\omega}{d\theta}\boldsymbol{\psi}_{\mathrm{ph}}>}{<\boldsymbol{\psi}^{\dag}_{\mathrm{ph}}|\mathrm{B}\boldsymbol{\psi}_{\mathrm{ph}}>}\,.
\end{equation}
 That is, the multistability of CSs caused by time-delayed feedback is related to the presence of the local saddle-node phase bifurcations of the localized CS solution. 

\paragraph*{Delay-induced dynamics}
In order to analyze the delay-induced dynamical behavior of a single CS, one-dimensional direct numerical integration of the system (\ref{eq:model1})-(\ref{eq:model3}) has been performed using the pseudospectral method for spatial derivatives on an equidistant mesh combined with the classical Runge-Kutta time-stepping. As was mentioned above, increasing the product of the delay strength $\eta$ and delay time $\tau$ above the threshold given by Eq.~\eqref{eq:drift_phase_bif} leads to a drift pitchfork bifurcation when the stationary CS loses stability, giving rise to a pair of branches of CSs moving uniformly along the x-axis in opposite directions. Notice that since the rotation matrix $\mathrm{B}$ in Eq.~\eqref{eq:drift_phase_bif} depends on the feedback phase $\varphi$, the bifurcation threshold also depends on this phase. This means that similarly to the case of a broad-area semiconductor cavity operating below the lasing threshold and subjected to a coherent optical injection discussed in~\cite{PimenovPRA13} the drift bifurcation exists only in a certain interval of feedback phases. The region of the drift instability in the $(\varphi,\,\eta)$ plane calculated numerically for two delay times $\tau=70$ and $\tau=150$ is shown in Fig.~\ref{fig:ind_drift_bd}~(a),~(b) in light blue, whereas the dark blue stays for the region, where CSs are stable. 
\begin{figure}[!ht]
 \begin{tabular}{l}
  \includegraphics*[width=0.5\textwidth]{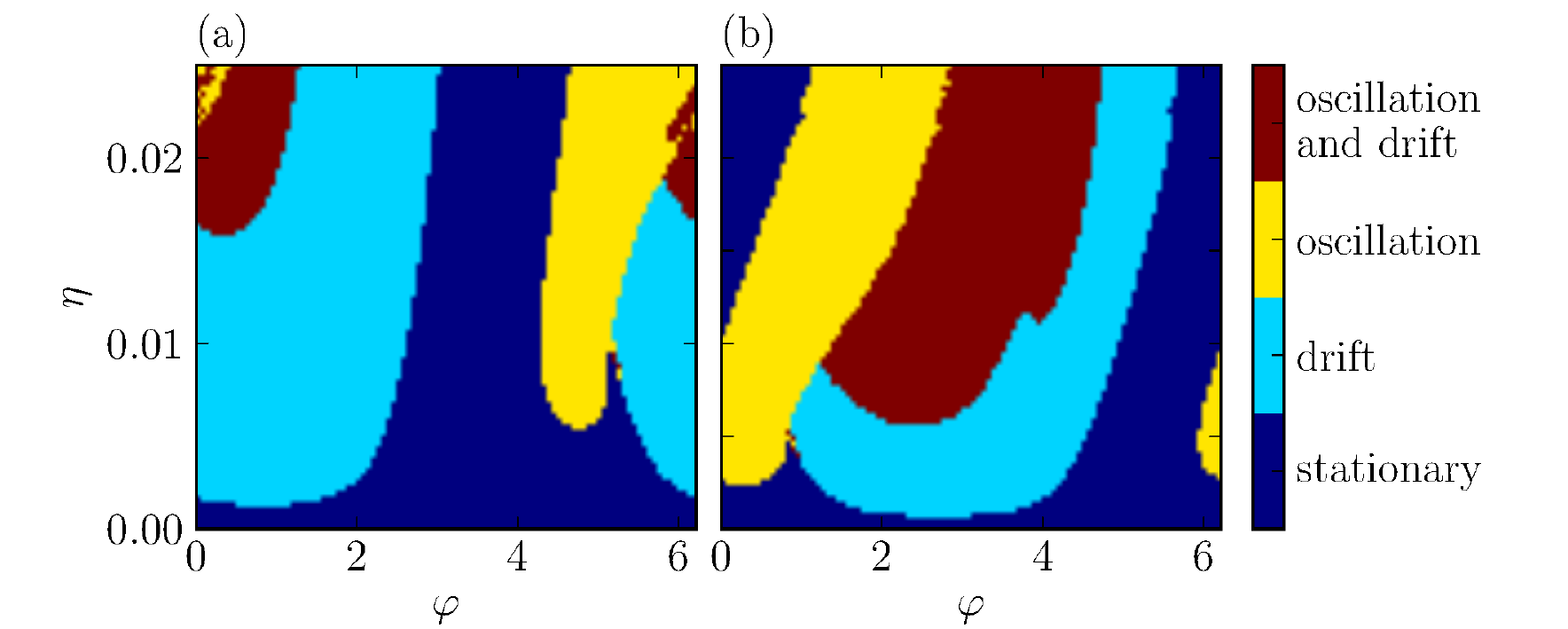}\\
  \includegraphics*[width=0.5\textwidth]{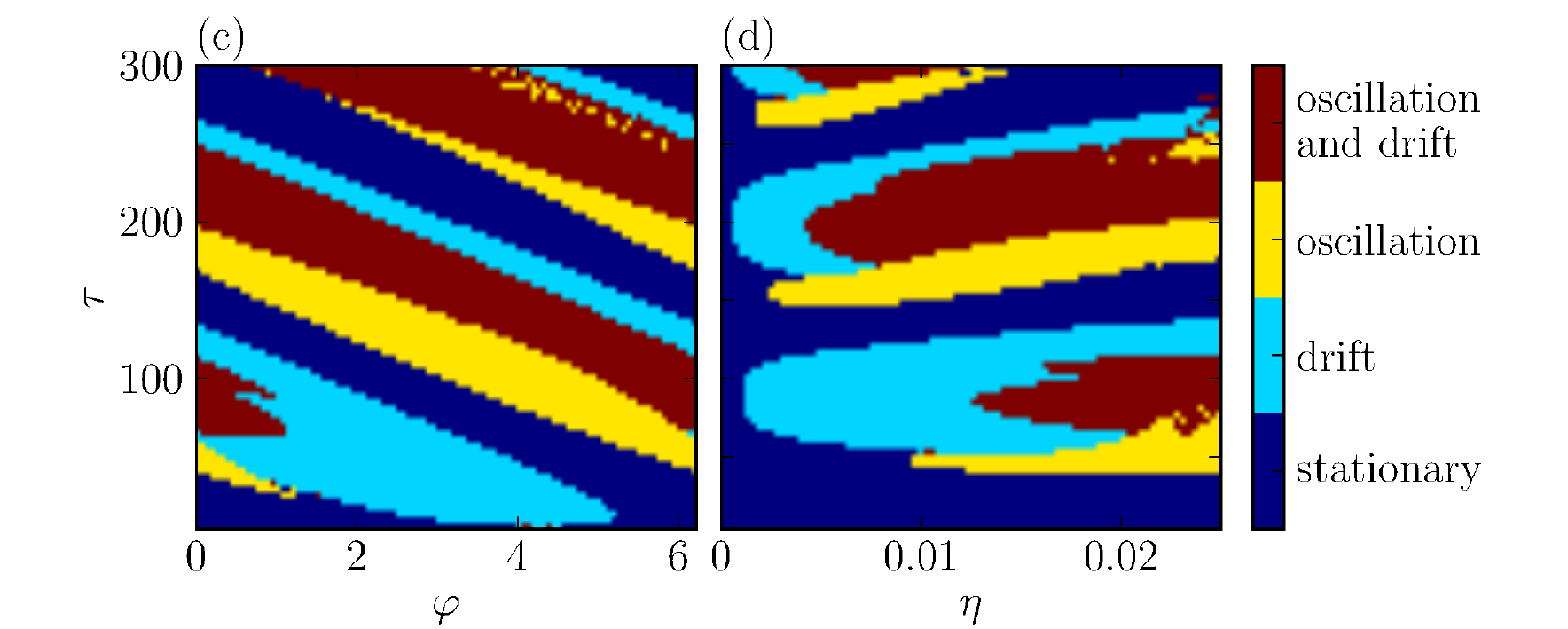}
 \end{tabular}
 \caption{(a), (b) Bifurcation diagrams for an one-dimensional single CS of the model~\eqref{eq:model1}--\eqref{eq:model3} in the $(\varphi,\,\eta)$ plane, calculated for the fixed value of the delay time (a) $\tau=70$, (b) $\tau=150$. (c), (d) Bifurcation diagrams in the $(\varphi,\, \tau)$ and $(\eta,\,\tau)$ planes calculated for the fixed values $\eta=0.02$ and $\varphi=0$, respectively. The region of stability of the CS is indicated in dark blue, whereas light blue, yellow and red regions correspond to a delay-induced drift, modulational and drift and modulational bifurcations, respectively. Other parameters are: $\alpha= 0.0, \beta= 0.0, b_1 = 0.9, b_2 = 1.0, \gamma = 2.0, s= 10.0, \mu = 2.06, d=0$.}
 \label{fig:ind_drift_bd}
\end{figure}
One can see that for both delay times the CS can become unstable with respect to the drift bifurcation at relatively small values of delay strength $\eta$. In addition, when increasing the delay time $\tau$ the region of the drifting CSs shifts in the direction of larger delay phases.

In addition to the drift bifurcation leading to traveling solutions and the phase bifurcation giving rise to the multistability of localized solutions, time-delayed feedback can also induce an Andronov-Hopf bifurcation.  In particular, in~\cite{Panajotov:14}, it was shown that the inclusion of a feedback term in the model~(\ref{eq:model1})-(\ref{eq:model3}) leads to the formation of breathing localized states as well as to a period doubling route to chaos. In addition, a delay-induced modulational instability can be observed in the limit of large delay times~\cite{PuzyrevPRA16}. Indeed, while the drift instability can be associated with the discrete eigenvalues, an instability of the pseudocontinuous part of the spectrum can induce a long-wavelength modulational instability, where a set of eigenvalues belonging to a branch of a pseudocontinuous spectrum with the translational zero eigenvalue at the origin becomes weakly unstable. At the modulational instability point the second derivative of this branch becomes positive in the origin~\cite{Wolfrum2010,Lichtner2011,Sieber20133109}. In~\cite{PuzyrevPRA16} the threshold for the modulational instability for the case of instantaneous gain and absorption relaxation was derived. It was shown that  the modulational instability gives rise to a sequence of Hopf bifurcations taking place above this threshold. Since the branch responsible for the modulational instability originates from a neutral translational eigenvalue, the corresponding eigenfunctions are complex, odd and have morphology similar to that of the translational Goldstone mode $\boldsymbol{\psi}_{\mathrm{tr}}$. That is, above the threshold of the modulational instability, the CS position starts to oscillate, giving rise to wiggling dynamics (see Fig.~\ref{fig:Zigzagging}~(a)). 
\begin{figure}[!ht]
 \includegraphics*[width=0.5\textwidth]{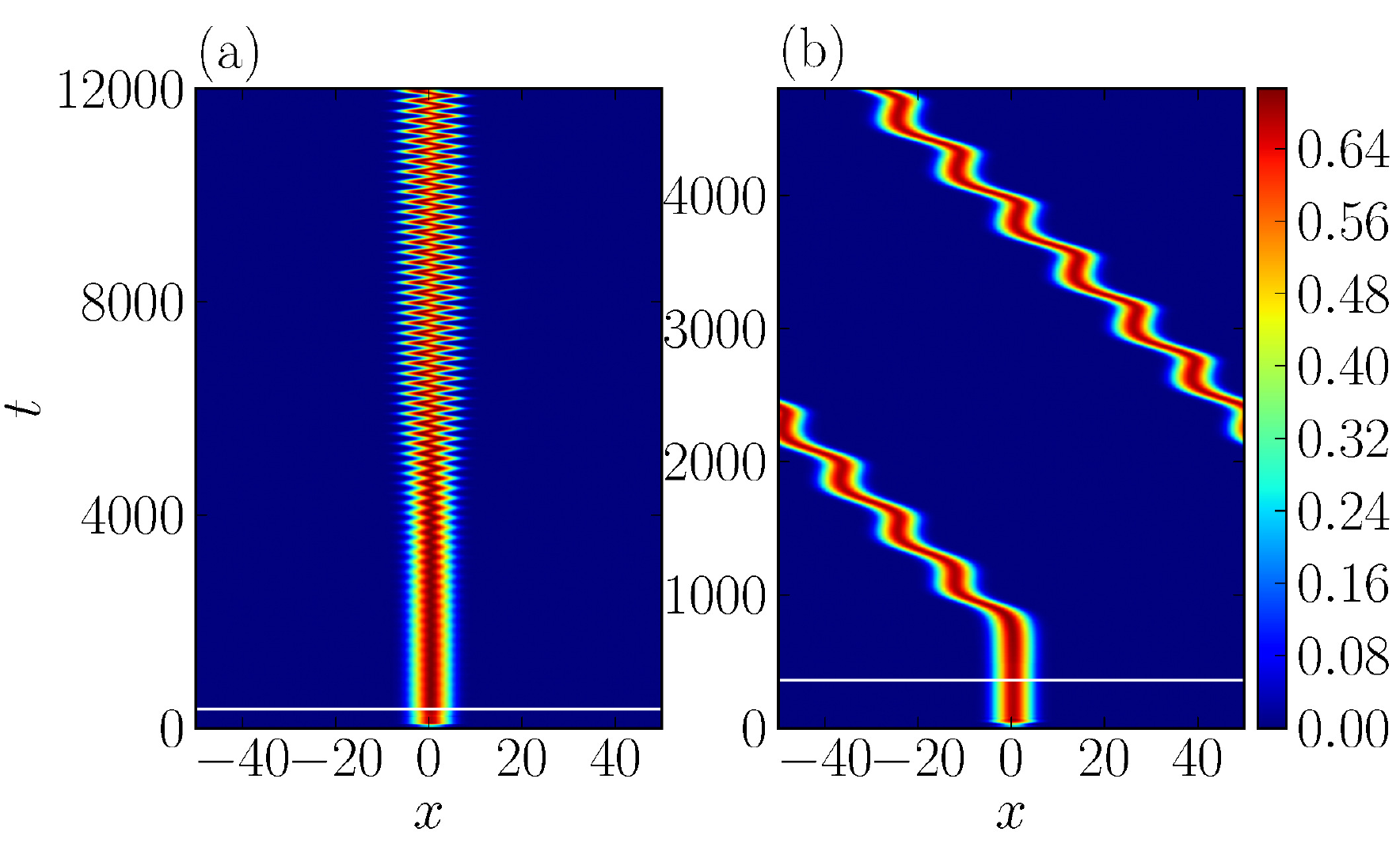}
 \caption{Space-time representation of the intensity field obtained by direct numerical simulation of Eqs.~\eqref{eq:model1}--\eqref{eq:model3} for $\varphi = 0$.  White line indicates the time step at which the time-delayed feedback is applied. (a)  Wiggling CS appearing above the modulational instability threshold for $ \tau = 150,  \,\, \eta = 0.0072$.  (b) Wiggling and drifting CS for $\tau = 320, \,\, \eta=0.005$. Other parameters are the same as in Fig.~\ref{fig:ind_drift_bd}.}
 \label{fig:Zigzagging}
\end{figure}
At larger delay times $\tau$, drift and modulational instabilities can interact leading to a formation of a moving and wiggling CS (see Fig.~\ref{fig:Zigzagging}~(b)). 

The regions in the $(\varphi,\,\eta)$ plane where CSs are modulationally unstable are indicated in Fig.~\ref{fig:ind_drift_bd}~(a),~(b) in yellow, whereas the domain of simultaneously wiggling and traveling CSs are shown in red. One can see that the regions of both instabilities move in the direction of smaller delay strength with increasing $\tau$.  CS bifurcation diagrams in $(\varphi,\,\tau)$ and $(\eta,\,\tau)$ planes calculated for the fixed values of $\eta$ and $\varphi$ are depicted in Fig.~\ref{fig:ind_drift_bd}~(c),~(d), respectively. In both panels one observes the bands of unstable regions, where instabilities leading to drift, wiggling or a combination of the two take place, separated by parameter regions where the CS remains stable.  The regions of drift become tiny with the increase of $\tau$, whereas the region of the modulation instability grows with $\tau$. Note that with increasing delay parameters, more and more complex eigenvalues from the pseudocontinuous spectrum may become unstable, leading to, e.g., a complex spatio-temporal behavior of the CS~\cite{PuzyrevPRA16}. 
\end{section}
\begin{section}{Stabilization of Spontaneous Motion}
As mentioned above, in the absence of the feedback term the Galilean symmetry of the system~\eqref{eq:model1}-\eqref{eq:model3} is broken for the finite relaxation rates $b_1$ and $b_2$. In general this leads to a shift of the corresponding ``Galilean'' eigenvalue from zero. As a result, for sufficiently large values of $b_1$ and $b_2$ a CS can loose its stability with respect to a drift bifurcation, giving rise to a formation of a SCS (cf. Fig.~\ref{fig:bdss} ). A control of the spontaneous motion of CSs with time-delayed feedback is a complicated issue and is still not understood to a large extent. In particular, for the Pyragas-like time delayed feedback term it was shown that a positive real eigenvalue generating a branch of delay-induced eigenvalues can never be stabilized by the feedback term, i.e., the corresponding eigenvalue can only approach the imaginary axis, but never cross it. Therefore,  stabilization of solutions with real positive eigenvalues is in general impossible with the chosen type of feedback force, whereas complex eigenvalues with positive real part can be stabilized~(see e.g., \cite{scholl2008handbook, Dahlem2008, KraftGurevich2016raey} for details).  In this subsection we show that an intrinsically moving SCS of  the system ~\eqref{eq:model1}--\eqref{eq:model3} can be stabilized by delayed feedback with the aid of a secondary drift bifurcation induced by the delay and discussed in the previous subsection.

\begin{figure}[!ht]
\begin{center}
\includegraphics*[width=0.5\textwidth]{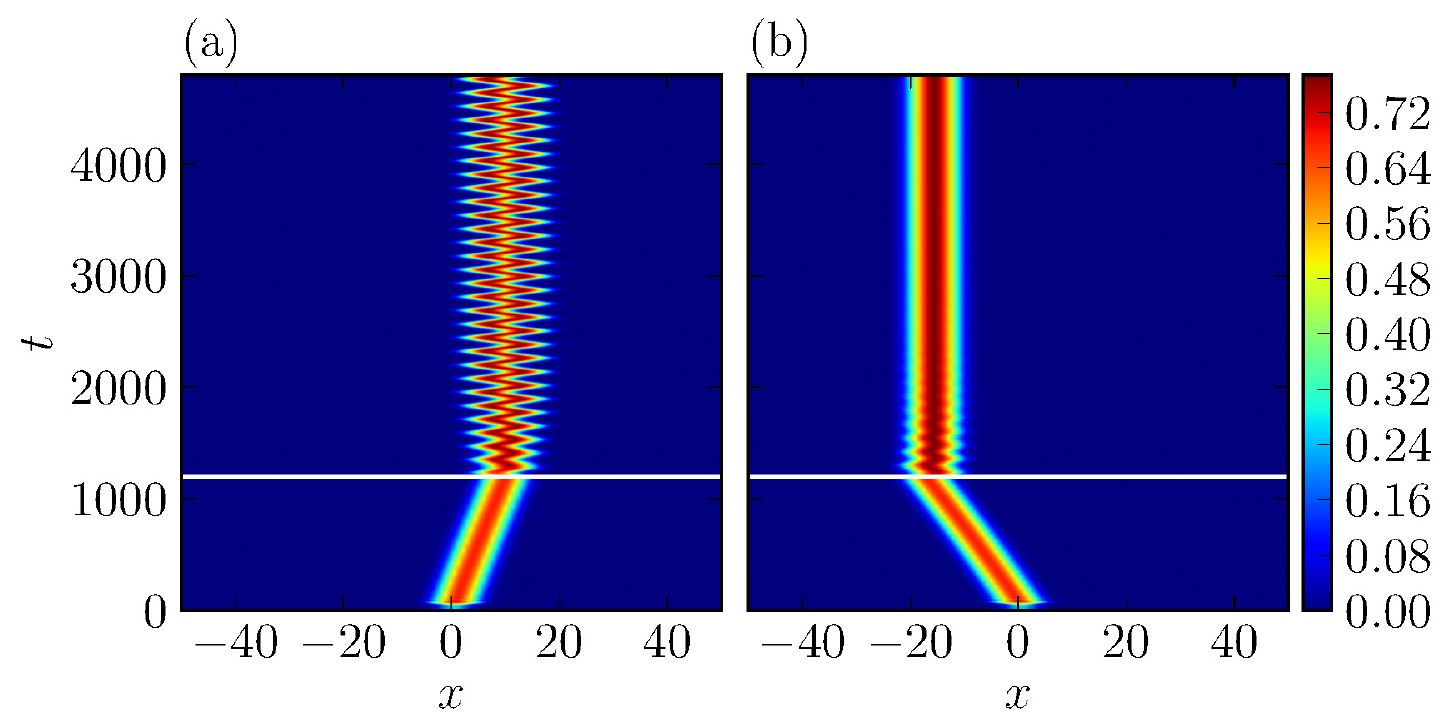}
  \caption{Stabilization of an intrinsically drifting CS. Space-time representation of the intensity field obtained by direct numerical simulations of Eqs.~\eqref{eq:model1}--\eqref{eq:model3} for $b_1^{-1} = 0.7\,, b_2^{-1}= 0.3$. After the feedback is turned on at some time indicated by white line for $\varphi=3.032$ the CS starts to follow a wiggling path of motion (a),  while for $\varphi=2.094$ the motion of the CS is suppressed (b).  The feedback parameters are $\tau = 100, \,\, \eta=0.02$. Other parameters are the same as in Fig.~\ref{fig:ind_drift_bd}.}
  \label{fig:stab_drift}
\end{center}
\end{figure}
Indeed, by choosing the feedback parameters to be in the region of the delay-induced drift (cf. Fig.~\ref{fig:ind_drift_bd}), one creates an additional branch of eigenvalues, generated by a real positive eigenvalue, responsible for a delay-induced drift. The corresponding eigenfunction is real, odd and has the same shape as the eigenfunction corresponding to the eigenvalue associated with an intrinsic drift. Hence, both real eigenvalues can collide and and leave the real axis at some feedback parameters. The resulting pair of complex conjugated eigenvalues can now be easily controlled by the feedback parameters. Figure~\ref{fig:stab_drift} shows an example of stabilization of a moving CS using the delay phase $\varphi$ as a control parameter. After the feedback term is turned on at the time moment indicated by a white line, an initially moving CS starts to oscillate around a stationary position (Fig.~\ref{fig:stab_drift}~(a)). By changing the phase $\varphi$, one can shift the pair of complex conjugated eigenvalues to the left half plane, suppressing the intrinsic motion completely (Fig.~\ref{fig:stab_drift}~(b)).

Figure~\ref{fig:stab_drift_bd} presents bifurcation diagrams of the intrinsically moving SCS in the $(\varphi,\, \eta)$ plane calculated using four different values of the delay time $\tau$. 
 \begin{figure}[!ht]
  \begin{tabular}{l}
   \includegraphics*[width=0.5\textwidth]{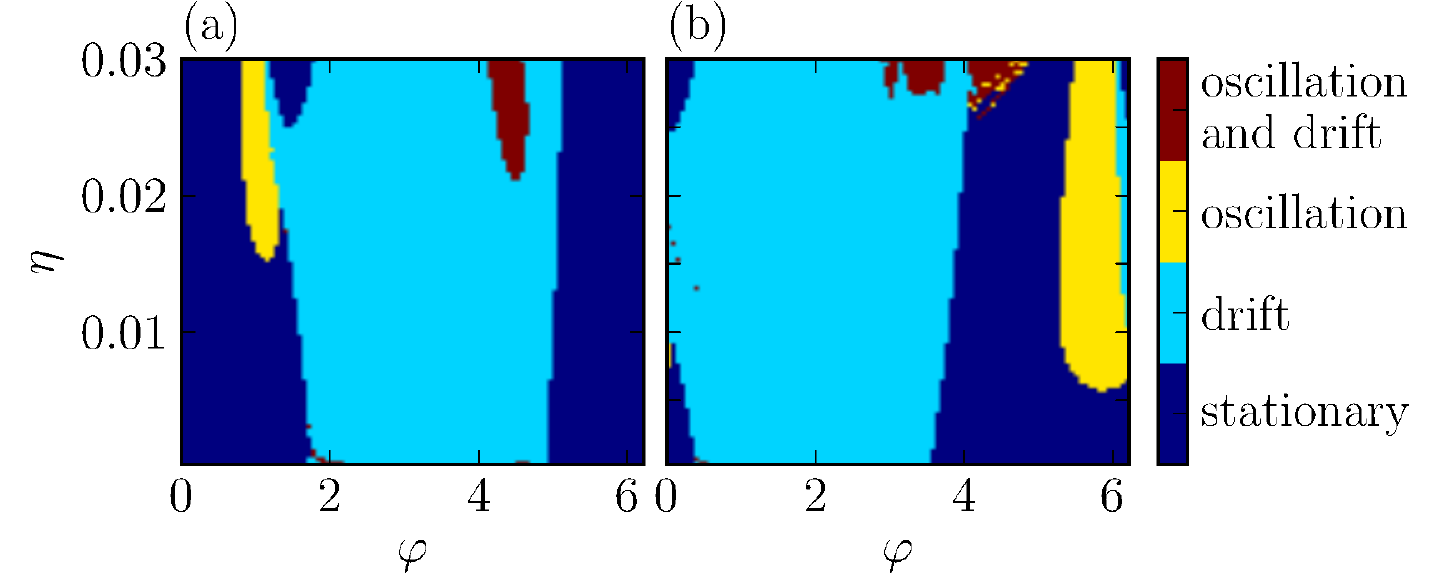}\\
   \includegraphics*[width=0.5\textwidth]{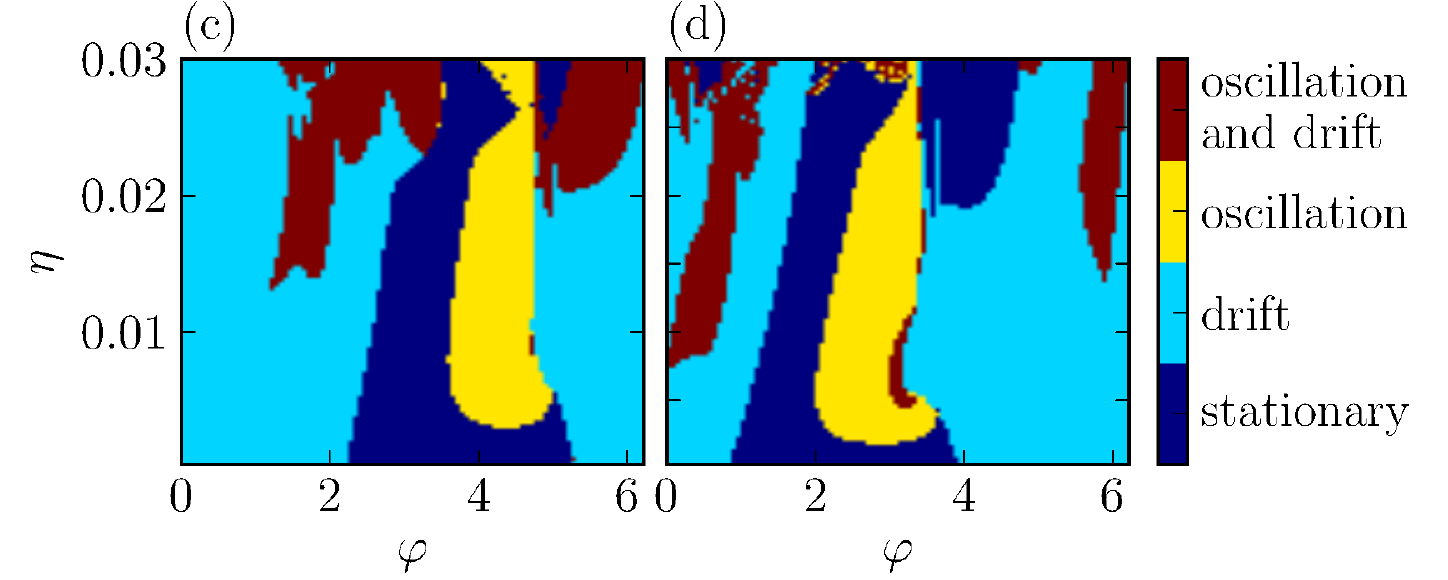}
 \end{tabular}
 \caption{Bifurcation diagram for an one-dimensional CS of the model~\eqref{eq:model1}--\eqref{eq:model3} in the $(\varphi,\,\eta)$ plane, calculated for the fixed values of the delay time (a) $\tau=25$, (b) $\tau=50$, (c) $\tau=75$, and (d) $\tau=100$. The regions where stabilzation of the intrinsic drift is successful are indicated in dark blue, whereas light blue, yellow, and red regions correspond to the double drift, wiggling, and drift and wiggling instabilities, respectively. Other parameters are the same as in Fig.~\ref{fig:ind_drift_bd}.}
\label{fig:stab_drift_bd}
 \end{figure}
 The regions where stabilization of the intrinsic drift is successful are indicated in dark blue, whereas light blue corresponds to the double drift bifurcation, where the stabilization fails and, depending on feedback parameters, SCS becomes either accelerated or slowed down. One can see that for relatively small delay times (see Fig.~\ref{fig:stab_drift_bd}~(a), (b)) the stabilization is successful in a wide range of feedback phases. The regions of wiggling oscillations (yellow) or wiggling and drift dynamics (red) are tiny and situated at large values of the delay strength $\eta$. However, for increasing delay times, these regions become more pronounced and are shifted in the direction of smaller values of $\eta$, while stable areas shrink. This can be explained by the fact that at sufficiently large $\tau$ a delay-induced modulational instability can take place, leading to complex oscillatory dynamics of the underlying system. 
\end{section}

\begin{section}{Delay-Induced Dynamics of Fast Cavity Solitons}

Finally, in this section we discuss the influence of the time-delayed feedback on the dynamics of the FCS. As mentioned in Sec.~\ref{sec:BACS}, FCSs are characterized by narrower intensity distribution and much larger velocities and peak intensities as compared to SCSs and they can coexist either with stationary or with slow moving CSs (cf. Fig~\ref{fig:bdss}).  An example of a transition from a SCS to a FCS is presented in Fig.~\ref{fig:fcs}~(a).
\begin{figure}[!ht]
\begin{center}
\includegraphics*[width=0.5\textwidth]{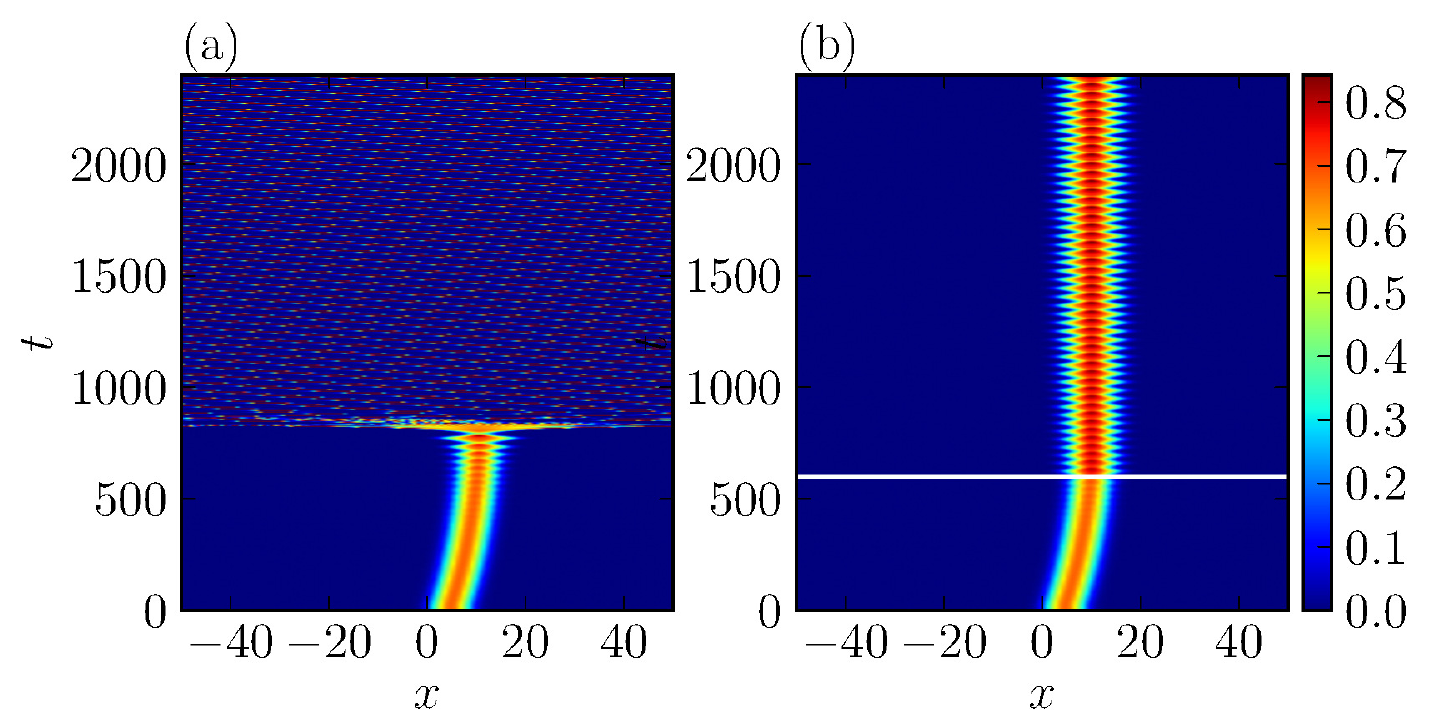}
 \caption{Left: In the absence of the delayed feedback, a SCS rapidly accelerates and a FCS is formed for $b_1=0.9$, $b_2=0.44$. Right: The same SCS transforms into a wiggling CS in the presence of time-delayed feedback. The transition to a FCS is arrested by the feedback. The feedback parameters are $\tau = 10,  \,\, \eta = 0.02,  \,\, \varphi = 0.0 $ (turned on at time $t=600$ and indicated by white line).}
 \label{fig:fcs}
\end{center}
\end{figure}
Numerical continuation performed in Sec.~\ref{sec:BACS} reveals that in the absence of time delayed feedback, the FCS bifurcates from the unstable part of a stationary CS in a subcritical pitchfork bifurcation and exists in a wide range of the gain parameter $\mu$. Due to this fact and since the velocity of the FCS is much larger than the velocity of the SCS, a direct control of the motion of the FCS is quite involved. However, one can arrest the transition to the FCS by applying to SCS time-delayed feedback. An example of such stabilization is shown in Fig.~\ref{fig:fcs}~(b). As it is seen from this figure, turning on the delayed feedback at the time, indicated by a white line, leads to a transformation of the SCS in to a breathing CS, so that the transition to the FCS is inhibited. The obtained breathing solution can be further controlled by changing the feedback parameters.

\begin{figure}[!ht]
\begin{center}
\includegraphics*[width=0.5\textwidth]{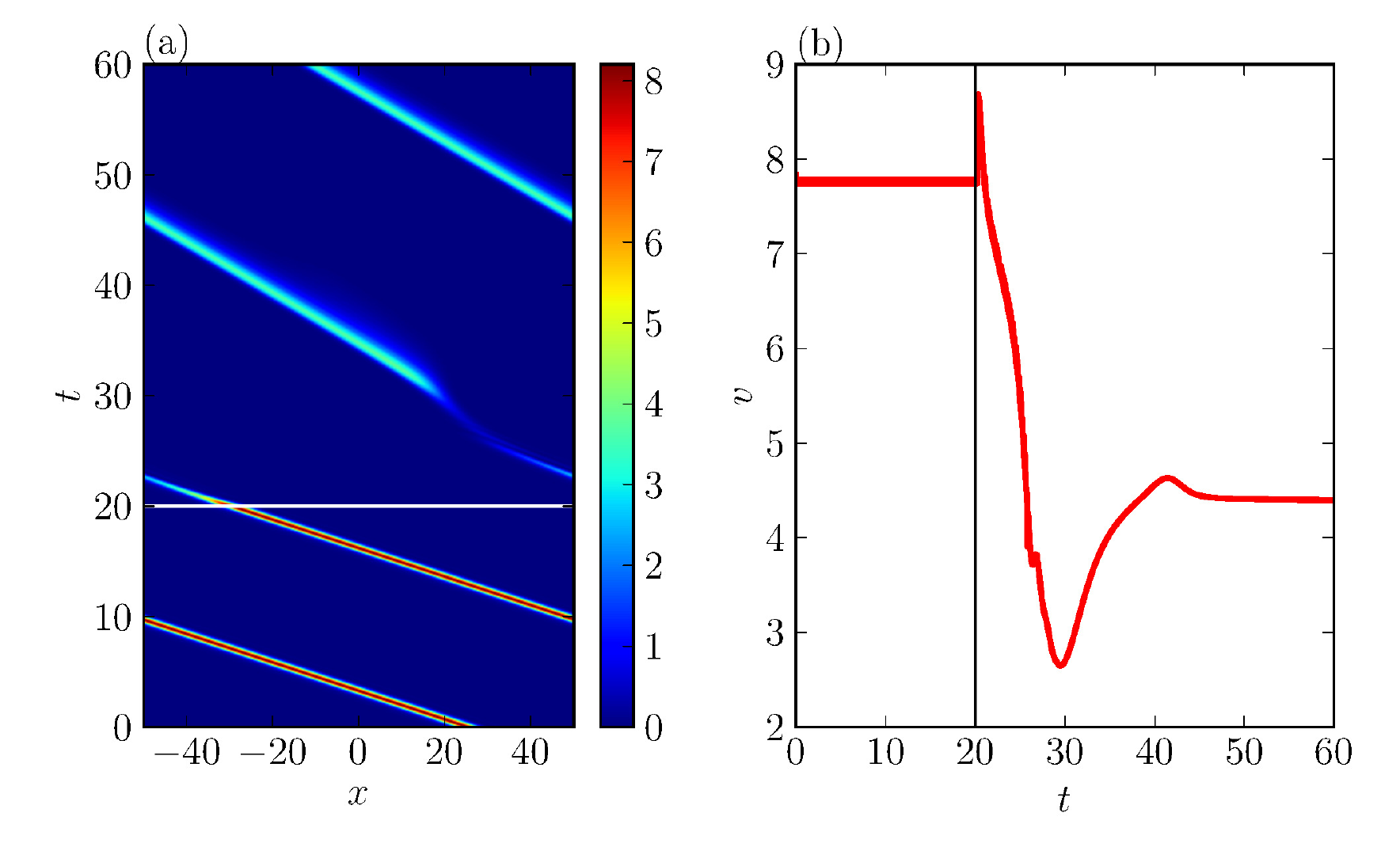}
 \caption{Space-time representation of the intensity field of the FCS (a) and its velocity as a function of time (b) obtained by direct numerical simulations of Eqs.~\eqref{eq:model1}--\eqref{eq:model3} for $b_1=0.34$, $b_2=1.0$. The FCS is slowed down by turning on delayed feedback at $t=20$ (white line). The feedback parameters are: $ \tau = 0.2,  \,\, \eta = 0.65,  \,\, \varphi = 1.113 $. }
 \label{fig:fcs_lower}
\end{center}
\end{figure} 
As mentioned in Sec.~\ref{sec:DelIndDyn}, the branches of stationary CSs form a tube shaped manifold in the presence of the delayed feedback. The branch of a single CS makes a number of turns around this tube giving rise to a multistability of CSs. Although the branches of slow and fast moving CSs are not easily computable for non-vanishing delays, one can expect that they exhibit the same snaking structure as stationary CS. That is, several stable and unstable moving CSs might coexist for some fixed parameter values. Since a FCS exists only as a moving object, complete stabilization of this particular solution is hardly possible. However, one can control the speed of the FCS by switching to another solution branch with a lower drift velocity. An example of such a transition is presented in Fig.~\ref{fig:fcs_lower}~(a). One can see that after turning on the feedback, the solution jumps from initial FCS to another FCS moving at much smaller velocity (see Fig.~\ref{fig:fcs_lower}~(b)).
\end{section}
\begin{section}{Conclusion}
In this paper the influence of delayed optical feedback on the dynamical properties and stability of CSs in a transverse section of a broad area VCSEL with saturable absorption has been discussed. In the absence of the delayed term, the branches of stationary and moving CSs have been analyzed in detail using numerical path continuation methods and the bifurcations responsible for the appearance, stabilization, and destabilizarion of two types of moving CSs, ``slow'' and ``fast'' ones, have been investigated. In particular, we have demonstrated that the branch of fast moving CSs originating from the unstable part of the stationary CS branch can remain stable above the linear lasing threshold where it might be interpreted as a self-starting lateral-mode-locking regime in a single-longitudinal-mode laser with one-dimensional transverse section. We have shown that the delayed optical feedback impacts both stationary and moving localized solutions. In particular, we have demonstrated how the delayed feedback induces drift and multistability of CSs. In addition, we have shown that apart from the drift and phase instabilities, a CS can exhibit a delay-induced modulational instability associated with the translational neutral mode, which in a combination with the drift instability can cause oscillatory dynamics and wiggling motion of the underlying solution. Furthermore,  time delayed feedback has proved itself to be an effective method to stabilize  intrinsic motion of CSs using a secondary delay-induced drift bifurcation.
\end{section}

\bigskip
F.T. thanks the Studienstiftung des deutschen Volkes for their financial support. A.P. and A.G.V. acknowledge the support of SFB 787, project B5 of the DFG. S.G. acknowledges the support of Center for Nonlinear Science (CeNoS) of the University of M\"{u}nster.  A.G.V. acknowledges the support of the Grant No. 14-41-00044 of the Russian Scientific Foundation.


%

\end{document}